\pdfoutput=1
\RequirePackage{ifpdf}
\ifpdf
\documentclass[pdftex]{sigma}
\else
\documentclass{sigma}
\fi

\numberwithin{equation}{section}

\usepackage{bbm}

\let\a=\alpha   \let\g=\gamma  \let\d=\delta
\let\th=\theta   
\let\m=\mu               \let\om=\omega
\let\s=\sigma  
 \let\eps=\epsilon

\newcommand{\f}{\frac}

\def\tl{\tilde}
\newcommand{\Id}{{\mathbbm 1}}
\newcommand{\p}{\partial}

\newcommand{\tr}{{\rm tr}}

\newcommand{\la}{\langle}
\newcommand{\ra}{\rangle}

\newcommand{\cC}{\mathcal{C}}
 \newcommand{\cM}{\mathcal{M}}

 \newcommand{\R}{\mathbb{R}}

\newcommand{\spatial}{\Delta_3}

\newcommand{\immirzi}{\gamma}

\newcommand{\cspacetime}{\mathcal{M}}
\newcommand{\cspace}{\Sigma}

\newcommand{\C}{{\mathbb C}}
\newcommand{\N}{{\mathbb N}}

\newcommand{\cA}{{\mathcal A}}

\newcommand{\cG}{{\mathcal G}}

\newcommand{\cK}{{\mathcal K}}

\newcommand{\cH}{{\mathcal H}}
\newcommand{\cN}{{\mathcal N}}

\newcommand{\SU}{\mathrm{SU}}
\newcommand{\SL}{\mathrm{SL}}

\newcommand{\SO}{\mathrm{SO}}
\newcommand{\U}{\mathrm{U}}

\newcommand{\Spin}{\mathrm{Spin}}

\newcommand{\su}{{\mathfrak{su}}}

\newcommand{\so}{{\mathfrak{so}}}
\newcommand{\spin}{{\mathfrak{spin}}}

\renewcommand{\sl}{{\mathfrak{sl}}}

\newcommand{\bra}[1]{\la {#1}|}
\newcommand{\ket}[1]{|{#1}\ra}

\newcommand{\vJ}{\vec{J}}
\newcommand{\vK}{\vec{K}}

\newcommand{\vsigma}{\vec{\sigma}}
\newcommand{\id}{\mathbb{I}}

\begin{document}

\allowdisplaybreaks

\renewcommand{\thefootnote}{$\star$}

\renewcommand{\PaperNumber}{052}

\FirstPageHeading

\ShortArticleName{Discrete Gravity Models and Loop Quantum Gravity: a Short Review}

\ArticleName{Discrete Gravity Models and Loop Quantum Gravity:\\ a Short Review\footnote{This
paper is a contribution to the Special Issue ``Loop Quantum Gravity and Cosmology''. The full collection is available at \href{http://www.emis.de/journals/SIGMA/LQGC.html}{http://www.emis.de/journals/SIGMA/LQGC.html}}}

\Author{Ma\"it\'e DUPUIS~$^\dag$, James P.~RYAN~$^\ddag$ and Simone SPEZIALE~$^\S$}

\AuthorNameForHeading{M.~Dupuis, J.P.~Ryan and S.~Speziale}

\Address{$^\dag$~Institute for Theoretical Physics III, University of Erlangen-N\"urnberg, Erlangen, Germany}
\EmailD{\href{mailto:dupuis@theorie3.physik.uni-erlangen.de}{dupuis@theorie3.physik.uni-erlangen.de}}

\Address{$^\ddag$~MPI f\"ur Gravitationsphysik, Am M\"uhlenberg 1, D-14476 Potsdam, Germany}
\EmailD{\href{mailto:james.ryan@aei.mpg.de}{james.ryan@aei.mpg.de}}

\Address{$^\S$~Centre de Physique Th\'eorique, CNRS-UMR 7332, Luminy Case 907, 13288 Marseille, France}
\EmailD{\href{mailto:simone.speziale@cpt.univ-mrs.fr}{simone.speziale@cpt.univ-mrs.fr}}

\ArticleDates{Received April 25, 2012, in f\/inal form August 06, 2012; Published online August 13, 2012}

\Abstract{We review the relation between Loop Quantum Gravity on a f\/ixed graph and discrete models of gravity. We compare Regge and twisted geometries, and discuss discrete actions based on twisted geometries and on the discretization of the Plebanski action. We discuss the role of discrete geometries in the spin foam formalism, with particular attention to the def\/inition of the simplicity constraints.}


\Classification{83C27; 83C45}

\renewcommand{\thefootnote}{\arabic{footnote}}
\setcounter{footnote}{0}

\section{Introduction}

The success of lattice gauge theories suggests that a discrete formulation of general relativity can play a major role in understanding the quantum theory.
A discretized path integral is indeed the starting point of approaches to quantum gravity such as quantum Regge calculus \cite{HamberReview} and (causal) dynamical triangulations \cite{LollReview11}. In both cases, general relativity is discretized using Regge calculus \cite{Regge}.
A useful alternative is to consider discrete actions based on connection variables. This has been considered in the literature \cite{CaselleDadda,Khatsymovsky89}, and it is one of the main rationales behind the construction of spin foam models \cite{PerezLR}.
It requires a suitable discretization of the connection variables, and in particular of the simplicity constraints needed to single out the metric degrees of freedom. The action of general relativity based on connection variables allows a reformulation of general relativity as a topological theory plus so-called ``simplicity constraints", which play an essential role. Our f\/irst goal is to review the various discretizations of the simplicity constraints which appeared in the literature.

One advantage of a discrete path integral based on connection variables is the possibility of interpreting its boundary states as the spin network states of Loop Quantum Gravity (LQG).
This brings to the foreground the question of f\/inding a discrete geometric interpretation for spin networks, a program started long ago by Immirzi \cite{Immirzi95discrete}, and f\/inally solved with the introduction of twisted geometries \cite{twigeo}, a suitable generalization of Regge geometries. Our second goal is to review the relation between LQG and these two dif\/ferent discrete geometries.

Regge geometries can be recovered from twisted geometries imposing suitable shape matching conditions, which guarantee the continuity of the piecewise-f\/lat metric. Such conditions are not present in the canonical formulation of LQG, although evidence exists that they are imposed dynamically, as we review below in Section~\ref{secSpinfoams}. It has been further argued in \cite{DittrichRyan} that these conditions can be naturally introduced in the canonical framework as a discretization of the secondary simplicity constraints. We will also review this proposal and the way these discretizations are used to def\/ine connection-variable based path integrals.

In an ef\/fort to organize the review logically rather than historically, we will focus f\/irst on the canonical theory, and leave the path integral for a later stage. We begin in Section~\ref{section2} with a brief  overview of Regge calculus, where the fundamental variable is the metric, and its  discretization furnished by the edge lengths of a triangulation of spacetime. This will allow us to appreciate the peculiarities of working with the connection as the fundamental variable. For instance, instead of edge lengths, one typically ends up with discretizations involving other geometric quantities, such as areas and angles.

Next, in Section~\ref{sec:kinematics} we review the relation between LQG on a f\/ixed graph and twisted geomet\-ries.
LQG is a continuous theory of quantum gravity, def\/ined as a projective limit/direct sum over graphs.
Truncating the theory to a given graph captures only a f\/inite number of degrees of freedom, and these in turn \emph{may} be used to describe a discretization of general relativity.
Indeed, from the viewpoint of LQG, there is a priori no need to interpret this set as discrete geometries. The usual description of the truncated Hilbert space involves in fact continuous, albeit f\/inite, degrees of freedom. This is the traditional interpretation of distributional holonomies and f\/luxes \cite{AshtekarReport, CarloBook,ThiemannBook}, and more recently an alternative but analogously continuous interpretation has been proposed in \cite{BianchiAB,FreidelGeiller}. On the other hand, it has been shown  that the same holonomies and f\/luxes describe certain discrete geometries, more general than the one used in Regge calculus, called twisted geometries \cite{twigeo}. They correspond to a collection of f\/lat polyhedra, which def\/ine in general \emph{discontinuous} piecewise f\/lat metrics and extrinsic curvature \cite{IoPoly, IoCarloGraph}.
In the special case of a triangulation, if one further imposes suitable shape-matching conditions, continuity of the metric is ensured and a Regge geometry is recovered. Imposing analogue shape-matching conditions on an arbitrary graph extends a notion of Regge geometry to arbitrary cellular decompositions. However, while the f\/irst can be described in terms of edge lengths, the latter must be described using areas and angles.
The resulting picture of a relation between spin networks and (the quantization of) discrete geometries has proved very useful to understand the spin foam dynamics, and found applications in dif\/ferent contexts such as calculations of $n$-point functions~\cite{npoint}, cosmological~\cite{cosmo} and black hole models~\cite{bh}.

In this initial part, there is no mentioning of simplicity constraints.
Indeed, we are dealing with ordinary $\SU(2)$ LQG, in which the simplicity constraints are already solved at the classical, continuum level. The constraints enter the picture if we consider a covariant version of LQG, in which the spin network states are based on the entire Lorentz group. In the rest of Section~\ref{sec:kinematics} we describe this formulation and how the simplicity constraints can be discretized. Their implementation leads to a notion of covariant twisted geometry, where the polyhedra have Lorentzian curvature among them.
This material paves the way for subsequent discussions concerning the path integral action.

In Section~\ref{secHolst} we review the construction of \cite{DittrichRyan}: One starts with the Holst action, and discretizes it in terms of holonomies and f\/luxes. The variables are parametrized in way motivated by Regge calculus.
The procedure allows to study the shape-matching condition as part of discretized secondary simplicity constraints, and perform a full reduction in which the shape-matching conditions are imposed, obtaining a def\/inition of Regge phase space, which has been further developed in \cite{Dittrich:2011ke, DittrichRyan2}.
The comparison of the approaches of Sections~\ref{sec:kinematics} and~\ref{secHolst} of\/fers a  deeper understanding of the relation between the space of discrete connections and Regge calculus, as well as a dif\/ferent perspective on the simplicity constraints.

Finally in Section~\ref{secSpinfoams} we review the role of discretized actions in constructing spin foam models.
This is a rapidly evolving research area, and we content ourselves with reviewing some of the main ideas and the dif\/ferent discretization schemes proposed in the literature. Emphasis is put on the role of the primary simplicity constraints, on the use of intuition from discrete geometries in the way they are realized in the spin foam path integral, and on the emergence of the shape matching conditions in the large spin limit.

 \section{Regge calculus}\label{section2}

A discrete version of general relativity was provided by Regge in 
\cite{Regge}. Spacetime is triangulated using a simplicial manifold $\Delta$ and, as fundamental metric variables, one assigns the lengths of all the edges, $\ell_e$:
\[
{\cal M}   \to   \Delta, \qquad g_{\mu\nu}   \to   \ell_e   .
\]
This assignment induces a piecewise-linear f\/lat metric on $\Delta$: each tetrahedral 3-cell is f\/lat along with its boundary triangles and edges. The curvature is all concentrated into the notion of a~def\/icit angle $\epsilon_t$ associated to each triangle $t$, and represents the failure of the sum of 4-dimensional dihedral angles at $t$ to equal $2\pi$:\footnote{A 4-dimensional dihedral angle $\theta_t^\s$ corresponds to the angle, at a~triangle $t$, between two tetrahedra within a~4-simplex $\s$.}
\begin{gather}\label{deficit}
\eps_t(\ell) = 2\pi - \sum_{\s\in t} \th_t^\s(\ell)   ,
\end{gather}
where $l$ denotes the set of edge lengths. It emerges that all aspects of this discrete geometry can be reconstructed from the edge lengths.  An $n$-dimensional dihedral angle $\th_t^\s$ constitutes the angle, at an $(n-2)$-dimensional hinge $t$, between two $(n-1)$-simplices $\tau_1$ and $\tau_2$, within an $n$-simplex $\s$:
\[
 \sin\theta_t^\s(\ell) = \f{n}{n-1}\f{V^{(n-2)}_t(\ell)   V_\sigma^{(n)}(\ell)}{V^{(n-1)}_{\tau_1}(\ell)   V^{(n-1)}_{\tau_2}(\ell)} .
\]
where we have written this formula in terms of the volumes of the various simplices involved. In turn, these volumes may be specif\/ied in terms of their associated Cayley matrices:
\begin{gather*}
(V^{(n)})^2=\frac{(-1)^{n+1}}{2^n(n!)^2}\det C^{(n)},
\qquad
C^{(n)} =
\begin{pmatrix}
0	& 1		& 1  			&  1 			&  \ldots   		&  1             \\
        	& 0  		& \ell_1^2     	&   \ell_2^2   	&\ldots     		& \ell_n^2       \\
         	&    		& 0                 	& \ell_{n+1}^2 	& \ldots 		& \ell_{2n-1}^2  \\
 	&      		&          		&   \ddots           	&     \ddots		& \vdots    \\
       	&      		&			&			&	0		& \ell_{(n^2-n)/2}^2\\
         &    		& 			&     			&  			& 0
\end{pmatrix}
  ,
\end{gather*}
where  $\{l_1, \dots, l_{(n^2-n)/2}\}$ is the subset of edges which constitute the $n$-simplex in question (and, incidentally, the Cayley matrices are symmetric).  If one further specif\/ies Cartesian coordinates on a 4-simplex $\s$, a f\/lat metric can be explicitly written for $\s$ as:
\begin{equation*}
 \d^{\mu\nu} = \f1{V^2}\sum_{e}\frac{\p V^2}{\p\ell_e^2} \ell_e^\mu \ell_e^\nu .
\end{equation*}

Coming to the dynamics, the action principle for Regge calculus is built through the direct discretization of the Ricci scalar in terms of def\/icit angles and reads:
\begin{gather}\label{SR}
S_R(\ell) = \sum_{t\in\Delta} A_t(\ell) \eps_t(\ell).
\end{gather}
If a boundary is present, then one needs the discrete equivalent of the Gibbons--Hawking bounda\-ry term.  This bounda\-ry term has the same form as \eqref{SR}, except that the $2\pi$ in the def\/inition~\eqref{deficit} replaced by $\pi$.
Like the continuum Einstein--Hilbert action, \eqref{SR} is unbounded. This stems from the following pair of inequalities: $2\pi[1 - n_4(t)/2] < \eps_t < 2\pi$, where $n_4(t)$ is the number of 4-simplices containing $t$. This implies that:
\[
2\pi k A_{\rm tot} < S < 2\pi A_{\rm tot},
\]
with $k=1-\max\limits_{t\in\Delta} \{n_4(t)\}/2$ and $A_{\rm tot} = \sum\limits_{t\in\Delta} A_t$. However, notice that one-sided boundedness can arise for special conf\/igurations.

The equations of motion are:
\[
\sum_{t\in e} \eps_t(\ell) \cot \alpha_e^t(\ell) = 0 .
\]
Here the sum is over triangles $t$ sharing the edge $e$ and $\alpha_e^t$ is the 2-dimensional dihedral angle within $t$, opposite to~$e$. The action and the corresponding Regge equations provide an approximation to general relativity that is accurate to second-order \cite{BrewinGentle}. To be more precise, one assigns initial data to the simplicial boundary such that it possesses a unique solution in the interior. Subsequently, one compares the obtained edge lengths $\ell_e$ with those given by the appropriate geodesics of the continuum solution. Analytic and numerical results show that the dif\/ference between the two goes like the square of the typical length, thus smoothly to zero in the continuum limit.

An important issue in the above argument concerns the symmetries of~\eqref{SR}. These have been studied in the literature \cite{DittrichDiffeoReview, RuthProgress97}, and are the object of a dedicated research plan by Dittrich and her group \cite{Bahr:2009ku,Bahr:2009qc,DittrichDiffeoLattice,Dittrich:2011ke} (see also~\cite{RovelliMagic}). We refer the reader to the recent review \cite{DittrichDiffeoLattice} and mention here only some minimal facts. The natural invariance under (active) dif\/feomorphisms of general relativity is destroyed by the discretization: generically, there are no displacements of the lengths that preserve the metric and only trivial relabelings remain. In this sense, the edge lengths are perfect gauge-invariant observables. However, a notion of gauge invariance can re-emerge in the form of (possibly local) isometries of the discrete metric. The typical example is the case in which the edge lengths describe a patch of f\/lat spacetime. In this case, the action is invariant under bounded vertex displacements preserving the f\/latness. This somewhat accidental symmetry actually plays an important role in assuring that we recover dif\/feomorphism invariance in the classical continuum limit: as one increases the number of simplices, while assuming that f\/ixed boundary data induce a unique classical solution with typical curvature scale, one hits a~point where the average curvature is approximately zero. Thus, vertex displacements are always a~symmetry in the continuum limit.

We conclude this quick overview of Regge calculus with some remarks, which will be useful to keep in mind while moving on.

\subsection{Area-angle Regge calculus}
Motivated by LQG and spin foams, one can consider taking the areas of the triangles as fundamental variables, instead of the edge lengths. This was proposed in \cite{Makela94, RovelliBasis} and some attempts have been pursued in the literature \cite{BarrettWilliams,Makela,Wainwright}. Notice that a generic triangulation has more triangles than edges and, even when the numbers match, the same area conf\/iguration can correspond to dif\/ferent edge sets \cite{BarrettWilliams}, thus constraints among the areas are needed to guarantee that a unique set of edge lengths is reconstructed.  The dif\/f\/iculty with this idea is that the required constraints are non-local with respect to the triangulation and no general form is known. A~solution to this problem has been found using a formulation in terms of areas and angles \cite{DittrichSpeziale}. Increasing the number of fundamental variables allows one to render the needed constraints in a~local manner. They can be written explicitly and the unique reconstruction of edge lengths proved. The constraints are of two types: the closure constraints, say $C_\tau$, local with respect to the tetrahedra, and the gluing (a.k.a.~shape-matching) constraints, say $C^\sigma_{ee'}$, local with respect to each pair of edges within a simplex $\sigma$. Using suitable Lagrange multipliers, the resulting action reads \cite{DittrichSpeziale}:
\begin{gather*}
S[A_{t},\phi_e^\tau,\lambda^\tau, \mu^\sigma_{ee'}] = \sum_{t} A_{t}   \eps_{t}(\phi) + \sum_\tau
\lambda^\tau \, C^\tau(A, \phi) +
\sum_\sigma \sum_{ee'\in \sigma} \mu^\sigma_{ee'}   {\cal C}^\sigma_{ee'}(\phi).
\end{gather*}

\subsection{On the choice of variables}

Taking the lengths as fundamental variables is very natural and due to the automatic rigidity of the simplices: specifying the edge lengths always specif\/ies a unique $n$-simplex. Furthermore, the formulae prescribing its geometry are quite simple as one can see from the above expressions.
There are however some drawbacks with this choice that become more evident when trying to quantize the theory. The f\/irst one is that the space of edge length conf\/igurations is much larger than the space of piecewise-linear f\/lat metrics. To ensure that one is really recovering a~Riemannian (or pseudo-Riemannian) metric, triangle inequalities need to be imposed. These guarantee the positivity of (space-like, for Lorentzian signature) simplicial volumes. While this might be simple to deal with in the classical setting, such conditions need to be additionally imposed in a path integral formulation, making it cumbersome to handle.

A second drawback is that the geometry is very rigidly Riemannian. There is, for instance, no room for torsion. On the other hand, a number of approaches to quantum gravity, including LQG, permit the presence of torsion, typically sourced by fermions. Modif\/ications of Regge calculus to include torsion have been considered in the literature \cite{CaselleDadda,Drummond:1986hb,Holm:1991fk,Schmidt:2001vd}.

\subsection{On the quantum theory}

The Regge action is taken as a starting point for both quantum Regge calculus \cite{HamberReview, RuthReview} and, when restricted to the sub case when all the edge lengths are the same (the relevant variables then become just the numbers of simplices), for (causal) dynamical triangulations \cite{LollReview11}. Both are path integral approaches and have obtained quite interesting results, including evidence for the existence of a continuum limit. On the other hand, the dynamical content of such a continuum limit is still insuf\/f\/iciently known, which partly motivates the search for alternative discretization schemes. Two specif\/ic dif\/f\/iculties of these approaches, related to the choice of variables, are the following. The f\/irst is to construct a Hilbert space for the boundary states of the path integral. The second concerns the unwieldy positive-metric conditions along with the ambiguities of the path integral measure. See e.g.~\cite{Dittrich:2011vz, RuthMeasure} on the measure for quantum Regge calculus.

A possible answer to the above questions is provided by the discrete geometric interpretation of LQG on a f\/ixed triangulation. As we show below, the alternative description provided by LQG based on connection variables can dispense with the triangle inequalities by implementing them automatically, it allows for torsion, it has a well-def\/ined Hilbert space and a prescription for the path integral measure.

\section{Canonical LQG and discrete geometries}
\label{sec:kinematics}

Historically, Loop Quantum Gravity and discrete Regge calculus remained somewhat detached from each other, despite several superf\/icial similiarities.  Ultimately, spin foam models facilitated the development of a precise link between the two~-- since they both provide a dynamics for the LQG on a f\/ixed triangulation {\it and} approximate, in the large spin limit, exponentials of the Regge action. A preliminary, yet incomplete link was f\/irst established for the Barrett--Crane model, while a more robust result holds for the EPRL model~\cite{BarrettLorAsymp}. In the last few years, it has been shown that the connection can be established already at the kinematical level \cite{IoPoly, twigeo,IoCarloGraph}. The existence of a connection between LQG and Regge calculus had been envisaged long ago \cite{Khatsymovsky89,Smolin89Regge} and in particular in the work of Immirzi \cite{Immirzi95discrete,Immirzi96NPPS, Immirzi94}. There is however one catch with respect to earlier expectations: Regge calculus is too \emph{rigid} to accommodate for all the degrees of freedom carried by a spin network. The precise correspondence between LQG on a f\/ixed graph and discrete geometries requires a generalization of Regge geometries.

To understand the nature of this generalization, let us f\/irst brief\/ly review the structure of LQG on a f\/ixed graph. In Ashtekar--Barbero variables, the phase space of general relativity is described by the conjugated pair:
$\{A^i_a(x), E^b_j(y)\} = \g \d^i_j \d^b_a \d^{(3)}(x,y)$.
The key ingredient of LQG is a specif\/ic smearing of this algebra \cite{Ashtekar:1998ak,FreidelGeiller,twigeo,ThiemannBook}. One introduces an oriented graph and its dual. Then, the connection is integrated along the links of the graph, so as to obtain a~notion of parallel transport, or holonomy, $g_l={\cal P} \exp \int_l A$. The conjugate f\/ield, the densitized triad, is smeared along surfaces of the dual graph. However, the correct smearing depends on the connection, in order to ensure that the f\/inal variable transforms nicely under gauge transformations. We denote this covariant f\/lux by $X_l$, and refer the reader to the cited literature for details. The fact that the smeared triad f\/ield depends also on the connection means that it does not commute with itself any longer. In fact, the Poisson brackets among the smeared variables on a given link are identical to those describing the phase space of a rotor, $T^*\SU(2)\simeq {\mathbb R}^3\times \SU(2)$, parametrized by the pair $(g,X)$. In particular, $X$ acts as a right-invariant vector f\/ield and its conjugated cousin $\tilde X=-g^{-1} X g$, which is associated with inverting the orientation of the graph, acts as the left-invariant one. Therefore, the smearing process has introduced a notion of discrete phase space $P_\Gamma$, associated with the graph, and given by the Cartesian product $P_\Gamma = \times_l T^*\SU(2)$. This phase space can be quantized via the familiar representation of the holonomy-f\/lux algebra on the Hilbert space ${\cal H}_\Gamma = L_2[\SU(2)^L, d\m_{\rm Haar}]$.

In the rest of this section, we shall review the interpretation of $P_\Gamma$ in terms of discrete geometries. But f\/irst of all, we stress that such an interpretation is a choice; it is not mandatory for the physical understanding of the theory. Indeed, what is usually done, and perfectly legitimate, is to view the classical degrees of freedom described by $P_\Gamma$ in terms of continuous, albeit singular, conf\/igurations, where the metric is zero outside of the graph and its dual. Recently~\cite{BianchiAB, FreidelGeiller}, it has been shown that there exist a somewhat ``dual'' continuous interpretation, where the connection is zero outside of the graph, and the triad is piecewise f\/lat, although not piecewise-linear f\/lat\footnote{Namely, this continuous interpretation can be visualized as a Regge-like geometry in which the links are not straight, but arbitrarily curved.}. On the other hand, it is possible to interpret the space in terms of discrete geometries. This interpretation has proved instrumental in many aspects of LQG, especially for the understanding of the spin-foam dynamics, but also for cosmological and black hole aspects.

\subsection{Twisted geometries}
\label{ssec:twisted}

The truncation of the theory to a f\/ixed graph can be interpreted as a discrete geometry. The necessary step to do so is a dif\/ferent parametrization of $P_\Gamma$. On each link $l$, we trade the holonomy-f\/lux pair $(g,X)$ for two unit vectors in $\R^3$, $N$ and $\tl N$, plus a real number $\rho$ and an angle $\xi$:
\begin{gather}\label{tgpar}
(X,g) \mapsto (N, \tl N, \rho, \xi)  : \qquad X = \rho N, \qquad g = n  e^{\xi\tau_3}  \eps  \tl n^{-1}.
\end{gather}
Here $n\equiv n(N)$ is a Hopf section for the bundle $\SU(2)\simeq S^2\times S^1$, thus $N=n\tau_3 n^{-1}$, $\tau_i$ are the $\SU(2)$ generators in the fundamental representation, with $\eps=2\tau_2$, and we have been using implicitly the standard isomorphism between the $\mathfrak{su(2)}$ algebra and ${\mathbb R}^3$. See \cite{twigeo} for more details. The interpretation of the new variables is the following. Each 3-cell is taken to be f\/lat, and equipped with an (arbitrary) orthonormal reference system.  The two quantities $N_l$ and $\tilde N_l$ are interpreted as the two (normalized) normals to the face (dual to) $l$, in the two reference frames associated to the two cells bounded by $l$. The quantity $|\rho_l|$ is the area of the face $l$ and the quantity $\xi_l$ is related to the extrinsic curvature of the complex at $l$. Such areas and normals def\/ine a certain notion of discrete geometry, as we now review.

As a f\/irst step, recall that one is interested in gauge-invariant quantities. That is, quantities satisfying the $\SU(2)$ Gauss constraint:
\begin{gather}\label{Gauss}
\cG_n = \sum_{l\ni n} X_l = 0 .
\end{gather}
where the subscript $n$ denotes a node of the spin network graph.  The constraints are f\/irst class and def\/ine gauge transformations on the constraint surface. Imposing \eqref{Gauss} at each node and dividing by its action gives the reduced, gauge-invariant phase space $S_\Gamma = P_\Gamma/\!/ \cG^N$, where $N$ is the number of nodes of the graph. When the closure condition~\eqref{Gauss}
holds, each conf\/iguration of areas and normals around a node def\/ines a \emph{unique} convex polyhedron (up to rotations and translation). This is the content of an old theorem due to Minkowski. We invite the interested reader to see \cite{IoPoly} for details on the space of shapes of polyhedra and on the explicit reconstruction procedure.

Next, the graph provides a notion of connectivity between the various polyhedra reconstructed in this way around each node. Gluing the polyhedra together is, however, non-trivial: in fact, adjacent polyhedra share a face with a unique area, $|\rho_l|$, but with a dif\/ferent \emph{shape}; the induced lengths and angles are in general dif\/ferent in the frames def\/ined by the two polyhedra. Hence, the metric reconstructed in this way can be discontinuous across the faces. In this sense, LQG def\/ines a more general notion of discrete geometry than the one used in Regge calculus. Part of this generalization is the possibility of using arbitrary cellular decompositions, and not just triangulations. This is a desirable feature. But the real novelty is the possibility of discontinuous metrics, which is a consequence of trading the link lengths for areas and normals\footnote{The fact that using areas as fundamental variables would have led to discontinuities was anticipated in~\cite{Wainwright}.}. Such discontinuity might appear appalling at f\/irst, but positive arguments can be made for this feature; after all, standard Regge calculus is torsion-free, whereas the kinematical phase space of Loop Quantum Gravity should carry room for torsion. In any case, shape-matching, or gluing, conditions that reduce a twisted geometry to a Regge geometry can be given explicitly, see \cite{DittrichSpeziale} for the case of a triangulation, and~\cite{IoPoly} for the general case. These are not present in the kinematical formulation of LQG, but appear to be automatically implemented in the semiclassical analysis of spin foams, as necessary conditions for the existence of the saddle point approximation, thus explaining the emergence of standard Regge calculus in the asymptotics.

In this geometric picture, the angles $\xi_l$ carry a notion of discrete extrinsic geometry among polyhedra. To clarify this point, consider f\/irst the special case when the shape-matching conditions hold, that is equivalent to say, the $(X,g)$ variables come from a Regge geometry. Let us further simplify things by choosing a common frame for the adjacient tetrahedra, so that we have a unique normal, say $N$. Then one can show~\cite{IoCarloGraph} that $g=e^{\g \th N}$, where $\g$ is the Immirzi parameter and $\th$ the usual dihedral angle. The general situation of independent frames can be realized through an additional contribution $g\to g \Gamma$ where $\Gamma \in \SU(2)$ is the rotation that rotates the f\/irst reference frame into the second.  Without loss of generality, we can parametrize $\Gamma = n\tilde n^{-1}_\alpha$, where $\tilde n_\alpha =  \tilde n(\tilde N) e^{\alpha\tau_3}$. Multiplying $g$ from the right by $\Gamma$ we get:
\begin{gather*}
g=n  e^{(\gamma\theta-\alpha)\tau_3}  \tilde n^{-1} .
\end{gather*}
Comparison with \eqref{tgpar} tells us that $\Gamma=n \tilde n^{-1}$ is the spin connection part, and $\xi = \g \th -\a$. Notice that $\a$ depends not only on the choice of section but also on the $\SU(2)$ frames in each cell. This shows up in explicit constructions, e.g.~\cite{Magliaro:2010qz}.

In the general case, without additional shape-matching conditions, $\xi$ still carries extrinsic geometry, but also the normals $N$ and $\tilde N$. Hence, the relation between $\xi$ and the usual dihedral angle becomes undetermined. Some of these aspects will be clarif\/ied further below in Section~\ref{secHolst}, where it is shown that prior to imposing the shape-matching conditions, the usual notion of dihedral angle associated to a triangle has an additional dependence on the side of the triangle, thus making it impossible to be directly associated to~$\xi$.

Summarizing, there exists a precise relation between spin networks and discrete geometries. A~spin network quantizes a classical space, $S_\Gamma$, whose points are usually parametrized as a~distributional assignment of holonomies and f\/luxes, but that equivalently def\/ine a collection of adjacent f\/lat polyhedra, with extrinsic geometry among them. From the viewpoint of the quantum theory, the relevance of the above construction concerns the interpretation of coherent states on a f\/ixed graph, which can then be visualized as a collection of polyhedra describing a twisted geometry. This interpretation of coherent states has found many applications in the literature, from facilitating $n$-point function calculations to the analysis cosmological models.

\subsection{Spinors and twistors}
\label{ssec:spin-twist}

A beautiful aspect of this description is that it can be derived from a much simpler system, that is an assignment of spinors, $(z^{s}_l,  z^{t}_l)$, on the source and target nodes of each link \cite{twigeo2}. Each spinor is then equipped with canonical Poisson brackets. The result is a phase space of spinors associated to the graph, $\times_l \C^4$, of dimension $8L$. Although not necessary for the geometric interpretation per se, this description will turn out to be useful below, when discussing the covariant version of discrete geometries. Hence, we brief\/ly introduce now the spinorial formalism, referring the reader to the lectures~\cite{IoZako} for details.

The relation between the spinorial and the twisted geometries/holonomy-f\/lux phase spaces is once again via a symplectic reduction: on each link, we introduce a scalar constraint imposing the matching of the norms of the two spinors:
\begin{gather}\label{He}
H_l = \la  z^{s}_l| z^{s}_l\ra-\la  z^{t}_l| z^{t}_l\ra = 0.
\end{gather}
We then have $\C^4/\!/ H \simeq T^*\SU(2)$ \cite{twigeo2}. The result is just a classical version of the Schwinger representation of the angular momentum. In the reduced phase space, the f\/luxes and holonomies are parametrized as follows\footnote{$|z^{s,t}] = \varsigma \ket{z^{s,t}}$ def\/ines the dual spinor, with $\varsigma$ the $\SU(2)$ complex structure.}:
\begin{gather}\label{defX}
X = \bra{z^s} \f{\sigma}2 \ket{z^s}, \qquad
g = \f{|z^s\ra[ z^t|-|z^s]\la z^t|}{\sqrt{\la z^s|z^s \ra \la z^t|z^t\ra}}.
\end{gather}
The closure condition \eqref{Gauss} reads:
\begin{gather*}
\cG_n = \sum_{l\ni n} \ket{z_l}\bra{z_l} - \f12{ \bra{z_l}z_l\ra} \Id = 0.
\end{gather*}
Therefore, the previously described, collection of polyhedra, which corresponds to a gauge-invariant spin network, descends from an assignment of a spinor on each half-link, satisfying the area matching conditions on the links and the closure conditions on nodes. The phase space structure of such a \emph{spinor network} is captured by the following action:
\begin{gather}\label{spinornet}
S_\Gamma[ z^{s,t}_l]
 \equiv
\int d\tau
\sum_l -i\la  z^{s,t}_l|\p_\tau z^{s,t}_l\ra
+\sum_l \Phi_l(\la  z^{s}_l| z^{s}_l\ra-\la  z^{t}_l| z^{t}_l\ra)
+\sum_v \sum_{l\ni v} \la  z^v_l|\Theta_v| z^v_l\ra,
\end{gather}
where the scalars $\Phi_l$ and the traceless matrices~$\Theta_v$ are the Lagrange multipliers for area-matching and closure constraints respectively. Finally, the two spinors on each link can also be interpreted in terms of a~twistor, with respect to which the area-matching condition~\eqref{He} imposes vanishing helicity. See the lectures~\cite{IoZako} and the original works~\cite{EteraSpinor,twigeo2,EteraTamboSpinor} for details and applications.

\subsection{Covariant theory and simplicity constraints}
\label{secCovariantSimplicity}

Thus far, we have dealt with canonical LQG in real variables.  We now wish to discuss the covariant picture, that is the initial Lorentzian phase space, and its reduction to the $\SU(2)$ one via the imposition of the simplicity constraints. The interest is to show that a relation to discrete geometries can be established already at the Lorentzian level, and that accordingly, the simplicity constraints acquire a geometric interpretation. The latter is consistent with the original construction by Barrett and Crane~\cite{BarrettCrane}, and places it into a more general framework, where, for instance, arbitrary cellular decompositions can be considered, and not just simplicial ones.

In the following, we assume that the reader possesses a certain familiarity with the fact that the theory can be formulated in covariant terms via the Holst action, and that the $\SU(2)$ holonomy-f\/lux algebra is the result of a specif\/ic phase space reduction. The basics of this reduction will be presented below in Section~\ref{secHolst} and full details appear in the parallel review~\cite{AlexandrovSigma}. For the sake of this section, all we need to keep in mind is the following: (i) the initial variables are a connection in the Lorentz algebra $\SL(2,\C)$, $\om^{IJ}$, and its conjugate f\/ield $\Pi_{IJ}$; (ii) the action gives rise to (primary and secondary) simplicity constraints, whose imposition (in a time-gauge-f\/ixed setting) gives the $\SU(2)$ theory. In particular, the primary simplicity constraints single out a metric from $\Pi$, whereas the secondary ones~-- which notably lead to the second class nature of the system -- impose compatibility between the metric and $\SL(2,\C)$ connection. After a canonical transformation, the reduced variables are conveniently described by the Ashtekar--Barbero connection.

Consider then the same smearing procedure as before, but applied this time to the covariant phase space: $\{\om^{IJ}_a(x), \Pi^b_{KL}(y)\} = \d^{IJ}_{KL} \d^b_a \d^{(3)}(x,y)$. Introducing this time $\SL(2,\C)$ holonomies $G_l$ and f\/luxes $J^{IJ}_l$ along the links, one is led to a classical phase space $P_\Gamma^{\rm cov}=\times_l T^*\SL(2,\C)$. Just to acquaint the reader with the latter, we recall that a quantization of the space and its natural Poisson algebra is realized by the Hilbert space $L_2[\SL(2,\C)^L, d\m_{\rm Haar}]$ and its representation of the $\SL(2,\C)$ holonomy-f\/lux algebra.  An orthonormal basis is given by covariant spin networks, which appear in the literature on \emph{covariant LQG} \cite{AlexandrovLivine} and projected spin networks \cite{EteraLifting, EteraProj}.

The space can be parametrized in terms of spinors and given an interpretation as covariant twisted geometries \cite{IoHolo,IoTwistorNet,WielandTwistors}. To do so, we need this time four spinors per link or alternatively two twistors. The initial space is thus $\times_l \C^8$, with spinors $(t^s, u^s, t^t, u^t)$ for the source and target nodes, equipped with canonical brackets. The parametrization of the covariant holonomy-f\/lux algebra in terms of the spinors reads:
\begin{gather}\label{defJK}
\vJ^L=\f12\la t^s |\vsigma|u^s \ra ,
\qquad
\vJ^R=\f12\la u^s |\vsigma|t^s \ra ,
\qquad
G = \f{|t^s\ra[ t^t|-|u^s]\la u^t|}{\sqrt{\la u^s|t^s \ra \la u^t|t^t\ra}}  .
\end{gather}
The left-right generators are related to rotations and boosts via $\vJ^{L,R} := (\vJ\pm i \vK)/2$. The generators built with the source/target spinors are the right/left-invariant vector f\/ields. One can also write down a group element decomposition like in \eqref{tgpar}, which now takes the form:
\begin{gather*}
G = n  T_\alpha  e^{\Xi\tau_3} \bar T_{\tl\alpha} \eps^{-1} \tl n^{-1} .
\end{gather*}
Here, the $n$ are the same Hopf sections as before, the  $T_\alpha$ matrices lie in the triangular subgroup of $\SL(2,\C)$, while $\Xi$ is once again an angle. We refer the reader to~\cite{IoTwistorNet} for details.

The constraint reducing $\C^8$ to $T^*\SL(2,\C)$ is a complex version of the area-matching condition:
\begin{gather}\label{cM1}
\cM = \bra{u}{t}\ra - \bra{\tilde{u}}{\tilde{t}}\ra ,
\end{gather}
which generate $U(1)^\C\cong\C$ shifts on the spinors leaving \eqref{defJK} invariant. On the reduced 12d space, a lengthy but simple computation \cite{IoTwistorNet} shows that the coordinates~\eqref{defJK} satisfy the Poisson-algebra of $T^*\SL(2,\C)$, with $J$ and $\tilde J$ again right- and left-invariant vector f\/ields, and $G$ in the def\/ining right-handed representation $\mathbf{(0,1/2)}$. By taking the hermitian conjugate $G^\dagger$, or alternatively by exchanging the spinors for their duals and vice-versa, one gets a left-handed representation $\mathbf{(1/2,0)}$.

Gauge-invariant quantities further satisfy the $\SL(2,\C)$ closure constraints:
\begin{gather*}
\sum_{l\in n}\vJ^L = \sum_{l\in n}\vJ^R = 0.
\end{gather*}
We def\/ine a \emph{twistor network} as the generalization to the Lorentzian case of a~spinor network: a~set of twistors, or bi-spinors $\ket{t_l^{s,t}}, \ket{u_l^{s,t}}$, satisfying both the matching \eqref{cM1} and $\SL(2,\C)$ closure constraints, and up to the corresponding $\C^L$ and $\SL(2,\C)^N$ transformations. The phase space structure of the gauge-invariant phase space, $S_\Gamma^{\rm cov}$, is thus captured by the following action:
\begin{gather}
S_\Gamma[t^{s,t}_l,u^{s,t}_l]
 =  \int d\tau
\sum_l -i\la u_l^{s,t}|\p_\tau t_l^{s,t}\ra -i\la t_l^{s,t}|\p_\tau u_l^{s,t}\ra \nonumber\\
\hphantom{S_\Gamma[t^{s,t}_l,u^{s,t}_l]=}{}
 +\sum_l \Phi_l (\la u_l^s|t_l^s\ra-\la u_l^t|t_l^t\ra)
+\sum_{v}\sum_{e\ni v} \la t^v_l|\Theta_v|u^v_l\ra.\label{actiontwi}
\end{gather}
A reduction to the previous $\SU(2)$ case is obtained if we identify the canonical $\SU(2)$ subgroup of unitary matrices, via $G^\dagger=G^{-1}$ and $\vec{J}{^L} = \vec{J}{^R}.$ This is achieved if we set $\ket{u}=\ket{t}:=\ket{z}$ and $\ket{\tilde u}=\ket{\tilde t}:=\ket{\tilde z}$. Then, \eqref{defJK} reduce to~(\ref{defX}), and the area matching~\eqref{cM1} to~\eqref{He}.

A geometric interpretation of the twistor networks is obtained doubling up the $\SU(2)$ picture of a collection of polyhedra. We now have a pair of spinors, $(u_l,t_l)$, for each face around a node $n$, and accordingly a bivector $J^{IJ}=(\vec{J}^L,\vec{J}^R)$ via \eqref{defJK}. The bivector represents the two-normal to the face embedded in Minkowski spacetime, in the frame of $n$. The chiral closure conditions together with Minkowski's theorem imply the existence of \emph{two} polyhedra, corresponding to the right- and left-handed sectors.

The geometric interpretation becomes more interesting if one includes the simplicity constraints. This turns out to identify the right- and left-handed polyhedra, and leads to a notion of \emph{covariant} twisted geometries, a collection of 3d polyhedra with arbitrary $\SL(2,\C)$ curvature among them. In the continuum theory, the simplicity constraints are unambiguous. See~\cite{AlexandrovSigma} for a~review, and also \cite{MikeLR,bimetric} for discussions. On the other hand, when working at the level of the truncated theory, one needs to adapt them to the graph. This procedure introduces ambiguities, and dif\/ferent realizations of the constraints are possible. In particular, much depends on the details of the variables used. Here we review what happens when using the covariant holonomy-f\/lux algebras, either in the standard parametrization or in the spinorial one. Dif\/ferent procedures exist in the literature, associated to other variables, and will be reviewed below in Sections~\ref{secHolst} and~\ref{secSpinfoams}. Consider f\/irst the primary simplicity constraints, that only involve algebra elements. Their role is to impose the conditions for the existence of a reconstruction theorem, by which the bivectors satisfying them and the closure constraint allow to reconstruct a (unique, up to global $\SL(2,\C)$ and translations) f\/lat 4-simplex in $\R^4$. This was shown initially in \cite{BaezBarrett,BarrettCraneLor,BarrettCrane}, and reviewed more recently in~\cite{BarrettLorAsymp, EPRlong}, where following~\cite{EPR}, a linear version of the initially quadratic constraints were used. Finally, a third form of the constraints has also appeared, which exploits the spinorial parametrization to achieve a holomorphic factorization.
We give a~formulation of the three possibilities in a way that applies to arbitrary cellular decompositions.

{\bf Quadratic constraints.} On the space of $\SL(2,\C)$ invariants, we have:
\begin{gather} \label{simplquadra}
\vJ^{R}_i\cdot\vJ^{R}_j=e^{2i\theta}  \vJ^{L}_i\cdot\vJ^{L}_j,\qquad\forall\, i,j ,
\qquad \g=\tan\f\theta2,
\end{gather}
where $\gamma$ is the Immirzi parameter. This form is closest to the continuum formulation, and it is the one originally used (with $\gamma=0$) in the Barrett--Crane model. An important aspect of \eqref{simplquadra} is that they do not Poisson-commute with themselves, since for both left and right sectors:
\begin{gather}\label{JJJ}
\{\vec{J}_i\cdot \vec{J}_j, \vec{J}_i\cdot \vec{J}_l \} =  \vec{J}_i \cdot \vec{J}_j \wedge \vec{J}_l .
\end{gather}
Therefore, although the initial continuum primary simplicity constraints Poisson-commute, their discrete counterpart does not so, and in fact, the brackets do not even close to form a genuine algebra. The discretization, and more precisely the non-commutativity of the f\/luxes, renders them a second class system.
There is only a subset that still Poisson-commute, given by the gauge-invariant, ``diagonal'' part of the constraints ($i=j$):
\[
\big(\vJ^{R}_i\big)^2 = e^{2i\theta} \big(\vJ^{L}_i\big)^2 ,
\]
or equivalently
\[
M_i = J^2_i-K^2_i +2\cot\theta   K_i\cdot J_i = 0 .
\]

{\bf Linear constraints.}
The linear form of the constraint was introduced in \cite{EPR}, and it states that the combination $(J_i-\g\star J_i)_{IJ}$ has a unique (timelike) normal for all $i$:
\begin{gather}\label{simplLin}
{\cal N}^I (J_i-\g\star J_i)_{IJ}=0 .
\end{gather}
This constraint is more familiar in the literature in the time gauge ${\cal N}^I=(1,0,0,0)$,
where it takes the form:
\begin{gather}\label{defC}
\vec C_i \equiv \vK_i +\gamma \vJ_i = 0 .
\end{gather}
Being linear, it reproduces only one of the two sets of solutions of the quadratic constraints. The second set is obtained inserting $\star$ in the scalar product, or equivalently through the f\/lip $\g\mapsto-1/\g$.

{\sloppy {\bf Holomorphic constraints.}
A third version of the constraints has been introduced in~\mbox{\cite{IoHolo, EteraHoloEucl}}. Observe f\/irst that both~\eqref{simplquadra} and \eqref{simplLin} can be immediately written in the spinor formalism using~\eqref{defJK}. If we do so, we notice that the linear constraints are quadratic in the spinor components, but are not holomorphic with respect to their natural complex structure, a fact which manifest itself in the non-commutativity~\eqref{JJJ}.  This suggests a solution to the long-standing issue of unclosedness of the simplicity constraints brackets, by seeking a new parametrization which realizes a holomorphic-antiholomorphic splitting. This can be achieved taking the following constraints, manifestly holomorphic:
\begin{gather}\label{simple_def}
\cC_{ij} \equiv [t_i|t_j \ra - e^{i\theta} [u_i|u_j \ra = 0 .
\end{gather}
As shown in \cite{IoHolo, EteraHoloEucl}, \eqref{simple_def} imply the quadratic constraints and thus also the linear ones. Their key property is that they Poisson-commute with each other:
\[
\{\cC_{ef}, \cC_{gh} \} =0 ,
\]
while of course $\{\cC_{ef}, \overline \cC_{gh} \} \neq 0.$
This is the key property of such holomorphic simplicity constraints, which has important applications at the quantum level. Notice that because of the Pl\"ucker relations, there are only $2N-3$ independent constraints per node. Nevertheless, they can all be imposed harmlessly since they commute.   Observe also that the distinction between diagonal and of\/f-diagonal constraints, familiar from the quadratic version \eqref{simplquadra}, now disappears\footnote{The distinction is however only apparent, because upon inclusion of the secondary constraints, the complete set is second class already in the continuum.}, with the advantage of a proper algebra and a clear holomorphic factorization.

}

The primary constraints can be added to \eqref{actiontwi}. It is natural to choose the holomorphic ones. We obtain a notion of \emph{simple} twistor networks, with action
\begin{gather}\label{actionsimple}
  S^{\rm simple}_\Gamma[t^{s,t}_l,u^{s,t}_l]
=S_\Gamma[t^{s,t}_l,u^{s,t}_l] + \int d\tau  \sum_n\sum_{l,\hat l\ni n} \Psi_{l,\hat l} \Big([t^n_l|t^n_{\hat l}\ra-e^{i\theta}[u^n_l|u^n_{\hat l}\ra\Big) ,
\end{gather}
with $\Psi_{l,\hat{l}} $ a suitable Lagrange multiplier.
The role of the simplicity constraints is then to identify the right- and left-handed sectors as in \eqref{simplLin}, up to a $\gamma$-dependent phase.
This identif\/ies a~unique polyhedron around each node, with the face bivectors all lying in the same 3d spacelike surface, plus a timelike normal, $N_n^I$, encoded in the spinors. The role of the Immirzi parameter is to determine the true area bivector as $B^{IJ}=(\id -\gamma\, \star) J^{IJ}$.
This information can be ef\/fectively traded for a \emph{single} spinor $ \ket{z^n_l} $ per half-link, and one pure boost $\Lambda_n \in \SL(2,\C)/\SU(2)$ per node, such that $\ket{t^n_l} = \Lambda_n \ket{z^n_l}$ and  $\ket{u^n_l} = (\Lambda_n)^{-1} \ket{z^n_l}$ \cite{IoHolo}.
In other words, a simple twistor network describes a~covariant twisted geometry: a collection of closed 3d polyhedra with arbitrary $\SL(2,\C)$ curvature among them.

Finally, there are also the secondary constraints. These include a relation between the densitized triad and the connection, which in the continuum implements the compatibility condition on the spatial slice. The solution provided by the Ashtekar--Barbero connection can be argued to be realized by the reduction $G=g$ of the group element. See \cite{AlexandrovNewVertex,IoCarloGraph}. Upon doing so, \eqref{actionsimple} reduces to \eqref{spinornet} and we immediately recover $\SU(2)$ LQG.
In support of this argument, notice that if we use the $\SL(2,\C)$ gauge invariance at the nodes, we can gauge-f\/ix all the boosts to the identity.  This also reduces \eqref{actionsimple} to \eqref{spinornet}. These two points highlight the pertinent aspects of the task at hand: the reduction to $\SU(2)$ LQG via the explicit solution of the second class constraint set. This second argument is somewhat reminiscent of the fact that second class constraints can be thought of to a certain extent as gauge-f\/ixing conditions.
More details on the reduction from $\SL(2,\C)$ to $\SU(2)$ holonomies and f\/luxes have recently appeared in~\cite{IoWolf}.

On the other hand, it has been argued in  \cite{AlexandrovNewVertex} that an alternative reduction should exist, amounting to the implementation at the discrete level the continuum solution proposed by Alexandrov. Such a reduction could be related to the discretization procedure which we review in the next section.

The simple twistor networks are very interesting from the perspective that they contain the same information as a normal spinor network for $\SU(2)$, but allow one to describe its natural embedding into an $\SL(2,\C)$-invariant structure, through the introduction of non-trivial time-normals living at each vertex of the graph $\Gamma$. They provide a classical version of the simple projected spin networks~\cite{IoHolo}, which form the boundary Hilbert space of EPRL/FK spin foam models~\cite{EteraLifting,IoCarloGraph}. In particular, even without the explicit implementation of the secondary constraints, these special class of spin network have the remarkable property of being entirely determined by their restriction to the $\SU(2)$ subgroup. See more on them below in Section~\ref{secSpinfoams}.

\subsection{Final remarks and the special case of 2+1 dimensions}

The above discussion has highlighted a key dif\/ference between LQG and Regge calculus. While truncating LQG on a f\/ixed graph can be interpreted as a~discretization of general relativity, this discretization is more general than Regge calculus, even when the graph is taken to be a~triangulation:  Holonomies and f\/luxes carry \emph{more} information than can be encoded in a Regge geometry.  This is not in contradiction with the fact that the Regge variables and the LQG variables on a f\/ixed graph both provide a truncation of general relativity: simply, they def\/ine two distinct truncations of the full theory. More details of this discrepancy will become clear in the next section, where we review a simplicial discretization of the Holst action in terms of a dif\/ferent parametrization of the holonomy-f\/lux variables, which allows a direct Regge interpretation of the simplicity constraints, and to see the emergence of Regge geometries if one further imposes the shape matching conditions.

But before going there, to better understand the physics behind these dif\/ferent truncations, let us brief\/ly consider a case where the dif\/ference disappears: this is what happens in 2+1 dimensions. In this case, the Regge picture and the connection picture coincide. The Regge picture is again described by an assignment of edge lengths, def\/ining a piecewise-linear 2d metric. Therefore, the conf\/iguration space on a given triangulation has dimensions $E$.
On the connection side, the (covariant) conf\/iguration space $\SU(2)^L/\SU(2)^N$ has dimensions  $3(L-N) = 3(F-\chi)$, where $\chi$ is the Euler characteristic of the surface. In turn, this graph is dual to the 2d triangulation, for which the relation $2E=3V$ holds. Hence, $3(L-N)=E$ which coincides with the Regge conf\/iguration space. Furthermore, $\SU(2)^L/\SU(2)^N$ itself is parametrized by the moduli space of f\/lat connections up to punctures. In other words, there is a~unique correspondence between a punctured Riemann surface and a piecewise-linear f\/lat metric, see e.g.~\cite{Carfora:2002rn}. A simple counting shows that this argument breaks down in higher dimensions, which is consistent with the result that the phase space of LQG is larger than Regge phase space.

\section{The Holst phase space:\\ continuum and discrete interpretations} \label{secHolst}

As mentioned earlier, Regge calculus is the prototypical example of a discrete geometrical theory. A~Regge geometry is a metric structure specif\/ied on a simplicial manifold. Moreover, quantum Regge calculus  has many of the desirable features that one would wish for a quantum theory of gravity: most notably, it def\/ines a quantum measure over discrete metrics. Of course, this def\/inition is not completely satisfactory, from a formal point of view, but even more so, for the limited evidence of a correct continuum limit (see however~\cite{Hamber:1996pj}). Among other things, explicit background independence is lost and it is dif\/f\/icult to quantify precisely the subsequent ef\/fects.
Note, however, that a basis for the boundary state space is labeled by abstract spin networks. These encode both the boundary (discrete) manifold and the dynamical degrees of freedom thereon.  These degrees of freedom inherited from a parameterization of the underlying classical phase space by 3d discrete metric geometries.   Importantly, this ref\/lects one common viewpoint in lattice quantizations: from the outset, they should mimic as closely as possible the character of the corresponding continuum theory.   Moreover, one might think of spin foam in this vein, that is, as a discrete quantization mechanism attempting to capture the physical character of gravity (and thus ameliorate certain negative aspects of quantum Regge calculus).  If one applies this philosophy to the Holst phase space of 3d continuum geometries, one would obtain a phase space of 3d discrete geometries, upon which one can reconstruct a metric.  In brief, one has once again the Regge geometry phase space.

On the other hand, spin foams may be viewed as a concrete implementation of the Hamiltonian constraint of Loop Quantum Gravity.  Loop quantum gravity is a canonical quantization mechanism starting from the Holst phase space.   Upon quantization (and implementation of all but the Hamiltonian constraint), one f\/inds that a basis for the boundary Hilbert space is given by embedded spin networks.   While these embedded spin networks capture certain aspects of the ambient smooth 3-manifold (within which they are embedded), the graph structure and the dynamical degrees of freedom are superf\/icially identical to those occurring in the discrete formulation above. Thus, the hope emerged that were one to dispense with the ambient 3-manifold, one could use techniques from the discrete theory to implement the Hamiltonian constraint, f\/ind the physical state space and def\/ine a suitable inner product. In other words, one could def\/ine a quantum gravity measure in this fashion.  We shall see later that modern spin foam models are remarkably successful in performing this task.

However, we have also spent some time in Section~\ref{sec:kinematics} describing the phase space at the root of the quantum states attached to a particular graph in Loop Quantum Gravity.   Interestingly, this phase space does not coincide with that of Regge geometries but with the rather larger one of twisted geometries. In this section, we shall review a discretization procedure that highlights how the shape matching conditions can be seen as part of the second class, secondary simplicity constraints.

\subsection{Some descriptive analysis}

Our modus operandi \cite{DittrichRyan, DittrichRyan2, Waelbroeck} is to discretize Holst's theory in its phase space formulation.  Thus, we discretize the spatial hypersurface along with the phase space variables, the symplectic structure and constraint set.

In passing from the continuum to the discrete, we set ourselves an initial task:  to devise a~discrete constraint set which reduces the initial phase space to that of twisted geometries; and to devise another that reduces it to that of Regge geometries.

\medskip

\noindent{\bf Result:} We f\/ind that both constraint sets are devisable, that the constraints have a similar functional form to their continuum counterparts and f\/inally that the twisted geometry constraints form a subset of the Regge geometry constraints.

\medskip

Thus we pass through the phase space of twisted geometries on our way to the phase space of Regge geometries.

An added boon of this discrete phase space approach is that we potentially have access to the symplectic structure on the reduced phase space via the Dirac bracket. The importance of the reduced symplectic structure
stems from very general considerations in path integral quantization, where it determines the quantum measure $\mu$:
\[
\mathcal{Z} = \int \mu_{\textrm{reduced phase space}} \, e^{\mathcal{-S}_{\textrm{reduced phase space}}} .
\]
Our second task was to compute this symplectic structure explicitly, both for the twisted geo\-met\-ry constraint set and for the Regge geometry constraint set.

\medskip

\noindent{\bf Results:}
\begin{itemize}\itemsep=0pt
\item For the twisted geometry constraint set, we found that the reduced symplectic structure coincided with that inspired by loop gravity \cite{twigeo}.
\item For the Regge geometry constraint set, we found that the reduced symplectic structure coincided with that inspired by Regge calculus \cite{Dittrich:2011ke}.
\end{itemize}

 Following on from this, one can give a def\/inition to two possible scenarios:
\[
\mathcal{Z}_{\textrm{twisted}} = \int \mu_{\textrm{twisted}} \, e^{\mathcal{-S}_{\textrm{twisted}}}\qquad
\textrm{or} \qquad
\mathcal{Z}_{\textrm{Regge}} = \int \mu_{\textrm{Regge}} \, e^{\mathcal{-S}_{\textrm{Regge}}}
 .
\]
A quantum dynamical theory of twisted geometries is realized by the modern spin foam models of Section \ref{secSpinfoams}, while quantum Regge calculus provides one possible quantum dynamics for Regge geometries.

Hidden inside these statements lies another interesting issue: the signif\/icance of the Immirzi parameter in such approaches to quantum gravity.  The twisted geometry symplectic structure is Immirzi parameter dependent, while the Regge geometry symplectic structure has no such dependence.  Thus, our analysis exposes the fact that we have an important decision to make in discretizing Holst's theory. On the one hand, if we are inspired by loop (quantum) gravity, we arrive at the phase space of twisted geometries and expect the Immirzi parameter to play a~signif\/icant role in our resultant quantum theory (at least at the outset).  On the other hand, if we are inspired to reproduce in our discrete theory the metric character of the continuum theory (at the outset), then we arrive at the phase space of Regge geometries and do not expect the Immirzi parameter to play any role in the quantum theory.

We note that the issue of Immirzi parameter signif\/icance has also raised its head in the continuum \cite{Alexandrov:2000jw, Geiller:2011bh, Geiller:2011cv}.  To a certain superf\/icial extent, they appear related.  In the continuum, as we shall see in a moment, one passes from the initial phase space to the phase space of 3d geometrical conf\/igurations by reducing with respect to a set of constraints (known as the simplicity constraints). One can deal with these simplicity constraints in two ways.  In the f\/irst, known as the loop gravity mechanism, one solves them explicitly.  In doing so, one arrives at a~theory on the reduced phase space that is Immirzi parameter dependent.  In the second, known as the covariant loop gravity mechanism, one constructs the reduced symplectic structure via the Dirac matrix.  This leads to a theory on the reduced phase space that is Immirzi parameter independent.  Classically, these two processes are totally equivalent, the results of which are just two parameterizations of the same theory. However,  as one can see, it raises questions as to the ultimate signif\/icance of the Immirzi parameter in the quantum theory.  Upon quantization, the Immirzi parameter labels a 1-parameter family of non-unitarily equivalent Loop Quantum Gravity theories and so plays an important role in the theory.  In fact, it also facilitates the very act of quantization.  Unfortunately, in passing to an Immirzi parameter free parameterization, covariant loop gravity renders the basic f\/ields highly non-commutative (classically) and thus makes quantization a highly-involved topic.

In comparing this to our discrete analysis, one might be tempted to view the Regge-to-Twisted geometries correspondence as the discrete analog of the covariant loop gravity-to-loop gravity relationship. There are certainly many similarities.  However, Regge and Twisted geometries are phase spaces of genuinely dif\/ferent sizes and characteristics.  Thus, we should not expect them to be related via a canonical transformation.  Rather they accentuate the point made in earlier sections that the Regge and Twisted geometry act as distinct (albeit related as we show below) truncations of general relativity.

We present now a very general description of the reduction process.

\subsection{Some methodological details}

We will consider the following action principle:
  \begin{gather} \label{eq:Holst}
S_{\textrm{Holst}}[e,\omega]=\int_\cspacetime  \tr_{\mathfrak{sl}(2,\mathbb{C})}\left( \star(e \wedge e)\wedge F[\omega] +\f1\gamma  (e\wedge e)\wedge F[\omega]\right) .
\end{gather}
The f\/irst term is just the Einstein--Cartan action for general relativity, to which it is equivalent assuming invertibility of the tetrad. The coupling constant $\gamma$ coincides with the Immirzi parameter \cite{Barbero, Holst, Immirzi96Real}. The term it multiplies is topological, and vanishes on-shell in the absence of torsion, thus making $\gamma$ classically irrelevant for pure gravity\footnote{It does play a role if a source of torsion is present, like fermions \cite{FreidelMinic, PerezRovelli}, although its ef\/fect is masked by non-minimal couplings \cite{AlexandrovFermions, Mercuri}.}. On the other hand, while one can only say that it {\it might} be relevant in the quantum gravity, it is certainly ubiquitous in LQG, where it enters the kinematical spectra of geometric operators as well as the covariant, spin foam description of the dynamics. Moreover, it has been recently shown to have a non-trivial renormalization f\/low \cite{IoDario,IoDarioProc, Daum:2010qt}.

To arrive at a phase space formulation, one performs a $3+1$ splitting of the spacetime manifold $\cspacetime = \R\times\cspace$.   We shall present here a brief digest of this structure for both continuum manifolds $\cspace$ and discrete manifolds $\spatial$:
\begin{gather*}\renewcommand{\arraystretch}{1.3}
\begin{array}{@{}r | c | c}
\hline
\hline
& \textsc{Continuum} & \textsc{Discrete} \\
\hline
\textsc{Spatial Manifold:} & \cspace & \spatial \\
\textsc{Phase space:} & (w(\vec{x}), \Pi(\vec{x}))  & (M_f, X_f)\\
\textsc{Symplectic structure:} &\{w,\Pi\}  = \delta & \begin{array}{@{}l}\{M_f,X_f\}  = C\,M_f\\[-0.3cm] \{X_f,X_f\} = C\,X_f  \end{array}\\
\textsc{Gauss:} & \mathcal{G}  & G\\
\textsc{Primary simplicity:}  & \mathcal{P} & P\\
\textsc{Secondary simplicity:} & \mathcal{S} & S_T \ \textrm{or} \ S_R ,\, S_T\subset S_R\\
 \hline
 \hline
\end{array}
\end{gather*}
For purposes of succinctness, we have suppressed all mathematical details.  However, let us explain the scheme of the above table.  In the continuum theory, the initial phase space is parameterized by the spatial components of an $\sl(2,\C)$ connection $w$ along with its conjugate momenta, which are certain components of an $\sl(2,\C)$ bi-vector f\/ield $\Pi$, at each point of $\cspace$.  They have the conventional symplectic structure, but they are subject to a system of constraints: the Gauss constraints $\mathcal{G}$ encode the local $\SL(2,\C)$ invariance;  the primary simplicity constraints $\mathcal{P}$ encode that $\Pi \sim e\wedge e$; while the secondary simplicity constraints $\mathcal{S}$ ensure that the primary simplicity constraints are preserved under evolution.  In fact, if one gauge-f\/ixes the boosts, one can give these simplicity constraints a more geometrical description.  In that case, the primary simplicity constraints force the bi-vector to be a spatial triad f\/ield; while the secondary simplicity constraints $\mathcal{S}$ ensure that the spatial connection is that one compatible with this triad.  We have left out the 4-dif\/feomorphism constraints as we shall not deal with them in the following.

\begin{figure}[t]
\centering
\includegraphics[scale = 1.3]{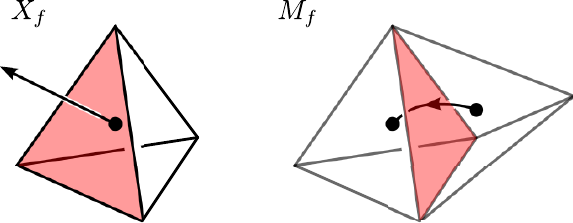}
\caption{Illustration of the discrete bi-vector and discrete connection.}\label{fig:variables-discrete}
\end{figure}

In the discrete theory, the initial phase space is parameterized by an $\SL(2,\C)$ matrix $M_f$ and an $\sl(2,\C)$ bi-vector $X_f$ attached to each triangle $f$ of $\spatial$, as illustrated in Fig.~\ref{fig:variables-discrete}.
 In particular, $M$ can be viewed as the parallel transport matrix mapping between the reference frames attached to each tetrahedron in $\spatial$.  One can arrive at the discrete symplectic structure via a rather transparent discretization method \cite{FreidelGeiller, Thiemann:2000bv}.  One f\/inds that it is the one compatible with the algebraic structure on $T^*\SL(2,\C)$, of which each pair $(M_f, X_f)$ form a representation.    The discrete Gauss constraints $G$ once again ensure $\SL(2,\C)$ invariance.  The discrete primary simplicity constraints $P$ ensure that we can construct a discrete metric geometry for each tetrahedron.  Meanwhile, the discrete secondary constraints $S_R$ ensure that the metric geometries constructed in adjacent tetrahedra are compatible at their intersection (that is, at their shared triangle).  One f\/inds that there is a subset of discrete secondary simplicity constraints $S_T\subset S_R$, which slightly relax this discrete metric compatibility and lead to the twisted geometries set out earlier.

\begin{figure}[htb]
 \begin{gather}\label{eq:variables-gauge}
 \renewcommand{\arraystretch}{1.3}
 \begin{array}{@{}r  |  c |  c  }
\hline\hline
\textsc{Variable}  &  \textsc{Label} & \textsc{Diagram}   \\ \hline
\textsc{Area} & A_f  & \includegraphics[scale=1]{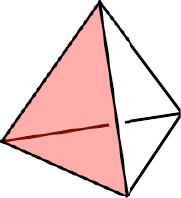} \\
\textsc{3d-dihedral angle} & \phi_e &  \includegraphics[scale=1]{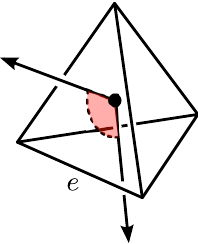}   \\
\textsc{4d-dihedral angle} &  \theta_{f}   & \includegraphics[scale=1]{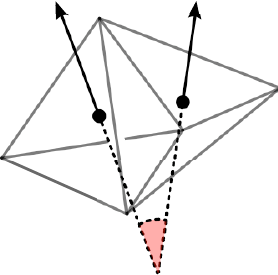}   \\
\hline\hline
\end{array}
\end{gather}
\end{figure}

{\bf Reduction type a: twisted geometry phase space.}
 Here, we reduce by the constraint set: $\mathcal{C}_T = \{G, P, S_T\}$.   Since we are reducing by the discrete Gauss constraints, we arrive at a~gauge-invariant phase space.  Thus, we found a convenient parameterization via gauge invariant quantities as exemplif\/ied in~\eqref{eq:variables-gauge}.

The reduced symplectic structure on the phase space is def\/ined via:
\[
\{\cdot,\cdot\}_T = \{\cdot,\cdot\} - \{\cdot,\mathcal{C}_T\}  \big[\{\mathcal{C}_T,\mathcal{C}_T\}\big]^{-1} \{\mathcal{C}_T,\cdot\}  ,
\]
where $\{\mathcal{C}_T,\mathcal{C}_T\}$ is the Dirac matrix.  We state its form here for the basis parameters:
\[
 \renewcommand{\arraystretch}{1.3}
\begin{array}{@{}c|ccc||c|ccc}
\hline\hline
\{\cdot,\cdot\}_T & A_f & \phi_e	& \theta_{f}
&
\{\cdot,\cdot\}_{\rm twisted} 	& A_f	&\phi_e		&\theta_{f}
\\ \hline
A_f		& 0		& 0		& (\star)
&
A_f		& 0		& 0		& \immirzi(\star)\\	

\phi_e	& 0		& \immirzi(\star)		&(\star)
&
\phi_e	& 0		& \immirzi(\star)		& \immirzi(\star)\\

\theta_{f}	& (\star)		&(\star)		& \frac{1}\immirzi(\star)
&
\theta_{f}		& \immirzi(\star) 	&\immirzi(\star)		& \immirzi(\star)\\
\hline\hline
 \end{array}
 \]
 In the above table, we have also described the corresponding symplectic structure $\{\cdot,\cdot\}_{\rm twisted}$ on the twisted geometries phase space.  To do so, we constructed analogous geometrical quantities from the twisted geometries basis.  To aid comparison, we have given them the same labels.  The marks $(\star)$ denote that corresponding entries have identical functional form.  As one can see, the only apparent dif\/ference is the precise power of the $\immirzi$-dependence. However, this dif\/ference stems from a rather subtle issue: how one deals with the $\SL(2,\C)$ Gauss constraint.  In our case, to arrive at $\{\cdot,\cdot\}_T$, we reduced by parameterizing the gauge orbits using $\SL(2,\C)$-invariant quantities.  To get to twisted geometries (via the loop gravity mechanism), one utilizes a 2-step approach.  One cuts the boost part of the orbits using a gauge-f\/ixing condition and then one parameterizes the rotation part using $\SU(2)$-invariant quantities.   One f\/inds that the $\theta_f$ constructed in the twisted geometries basis is af\/fected by boosting along the orbit (remember the dependence on the choice of section). Thus, the manifestly $\SL(2,\C)$-invariant def\/inition for~$\theta_f$ and the manifestly $\SU(2)$-invariant def\/inition for~$\theta_f$ are slightly dif\/ferent quantities.   Thus, we argue that the phase space, at which one arrives after reducing by $\mathcal{C}_T$, is identical to that of twisted geometries/loop gravity up to this dif\/ference in the choice of parameterization\footnote{Of course, in principle, one should be able construct a canonical transformation between the two parameterizations, but this is likely to be highly involved and we have not constructed it explicitly.}.

As a f\/inal remark, let us set out the tell-tale signs that this phase space does not describe Regge geometries. Obviously, it is larger than one would expect. Moreover, from the~$A_f$ and~$\phi_e$, one can reconstruct, within each tetrahedron, a set of six edge-lengths.  However, for two tetrahedra sharing an edge, one does not f\/ind that they assign it identical edge-lengths.  In the basis given above, this ambiguity manifests itself in the def\/inition of~$\theta_f$ (not the one we just mentioned but yet another).  Although an $\SL(2,\C)$-invariant quantity, $\theta_f$ is not truly the geometric 4d-dihedral angle.  (There are several non-equivalent $\SL(2,\C)$-invariant def\/initions on this phase space.)  To say it in yet another fashion, we do not possess a discrete spin connection.

{\bf Reduction type b: Regge geometry phase space.}
Here we reduce by the constraint set  $\mathcal{C}_R = \{G, P, S_R\}$. One f\/inds a particularly appropriate basis for the reduced phase space is given by the parameters $A_f$ and $\theta_f$.  The reduced symplectic structure on the phase space is def\/ined via:
\[
\{\cdot,\cdot\}_R = \{\cdot,\cdot\} - \{\cdot,\mathcal{C}_R\}  \big[\{\mathcal{C}_R,\mathcal{C}_R\}\big]^{-1} \{\mathcal{C}_R,\cdot\}
\]
and the only non-trivial commutation relation is:
\[
\{A_f,\theta_f\}_R = 1 .
\]
The areas and 4d-dihedral angles are canonically conjugate pairs.

We should remark on a number of issues at this stage.  First of all, for generic triangulations, the number of edges is less than the number of triangles.  Thus, it is in fact a subset of all pairs $(A_f,\theta_f)$ that parameterize the reduced phase space. On this phase space, one can show explicitly that the various def\/initions of the 4d-dihedral angle, which were inequivalent on the twisted geometries phase space, are now equivalent.  Another ef\/fect is that there is an unambiguous def\/inition for the length of each edge.  Finally, note the absence of the Immirzi parameter, the ramif\/ications of which we have commented extensively at the outset.

\section{Discretized actions and path integrals}
 \label{secSpinfoams}

Thus far, we have reviewed dif\/ferent aspects of discretizing connection variables for gravity,
and their relation to LQG and Regge calculus. In this section, we will review some aspects of these relations at the dynamical level.

\subsection{Hamiltonian formalism and recursion relations}

When talking about dynamics, the immediate problem concerns the fate of dif\/feomorphism symmetry. The experience with Regge calculus shows that the symmetry is broken, and can only be recovered in the continuum limit.
This can be shown at the classical level, but for the quantization the situation is much more complicated. Traditional approaches include quantum Regge calculus~\cite{HamberReview} and (causal) dynamical triangulations~\cite{LollReview11}. More recently, it has been proposed by Dittrich and collaborators to improve the Regge action in such a way as to make it carry an exact notion of dif\/feomorphism invariance.
Obtaining such a perfect action for 4d gravity is an extremely challenging problem, but progress might be achievable in a perturbative approach, e.g.~\cite{Bahr:2010cq,DittrichDiffeoLattice, Dittrich:2011vz}.

In a canonical formulation, the broken dif\/feomorphism symmetry leads to pseudo-constraints instead of proper constraints. Contrary to the latter, pseudo-constraints are proper equations of motion, in which the canonical data of two consecutive steps are (very weakly) coupled to each other \cite{Bahr:2009ku, Bahr:2009qc, DittrichDiffeoReview, Dittrich:2009fb, Gambini:2005sv, Gambini:2005vn}.
A priori, we see here an important tension with LQG, where an exact Hamiltonian constraint is always present.
The tension vanishes in the f\/lat case, because in this case exact constraints reappear in discrete gravity.
The exact form of such constraints can be found using the covariant Regge action as guiding principle.
Along these lines, in \cite{Bahr:2009ku, Dittrich:2011ke, Dittrich:2009fb} a canonical formalism for discretized gravity which exactly reproduces the dynamics as def\/ined by the discrete action was introduced. A discrete evolution scheme for Regge calculus is def\/ined using tent moves. Tent moves are a way of evolving locally a triangulated hypersurface such that the triangulation (that is the adjacency relations) of the resulting new hypersurfaces does not change \cite{Bahr:2009ku, Dittrich:2009fb}.
The consistency of these ideas with LQG restricted to a f\/ixed graph has been examined e.g.\ in~\cite{Bonzom:2011tf, Bonzom:2011hm, Bonzom:2011nv}. In~\cite{Bonzom:2011hm}, the authors built a new Hamiltonian for 3d gravity inspired from the f\/latness constraint on a triangulation in Regge calculus. In this context, the f\/latness equation becomes the statement that holonomies around the plaquettes $p$ are trivial. Projecting the curvature onto the components of the triad, they get an Hamiltonian, labeled by a given plaquette $p$ and a vertex $v$ in the cycle which bounds $p$. This Hamiltonian has a nice geometrical interpretation in terms of discrete geometries and dihedral angles of f\/lat simplices, and reads
 \[
H_{v,p}:= \sin \phi_{l_1 l}   \sin \phi_{l_2 l}  \left( \cos p_{t_1t_2}- \cos \theta_{t_1 t_2}\right).
 \]
In this expression, the $\phi$ and $\theta$ angles are 2d and 3d dihedral angles associated with a f\/lat tetrahedron, so functions of the lengths. $p$ are the momenta conjugated to the lengths. We see that the meaning of the Hamiltonian is to impose that the conjugated momenta coincide with the f\/lat dihedral angles, thus it asks for an embedding in f\/lat 3-space.
 The quantization of this Hamiltonian produces dif\/ference equations of order 2 when written in the spin network basis. The dif\/ference equations naturally come from the representation theory of the local group considered ($\SU(2)$ in the 3d case). On triangular plaquettes of the triangulation, the quantum equation $\hat{H}_{v,f}=0$ is the Biedenharn--Elliott identity, a recurrence relation which def\/ines the $6j$-symbol. The $6j$-symbol is also the Ponzano--Regge spinfoam amplitude and the Biedenharn--Elliott identity thus encodes the symmetry at the quantum level which makes the model topological.

 A similar situation can be realized also in 4d f\/lat models, such as BF theory \cite{Bonzom:2011tf}. As the $6j$-symbol is the physical state of BF theory on a tetrahedron (being in the kernel of the BF Hamiltonian), the $15j$-symbol is the physical state on a 4-simplex. Recursion relations for the $15j$-symbol were derived in \cite{Bonzom:2009zd} from the (regularized) 4-2 Pachner move. These recursion relations are dif\/ference equations contributing to the symmetries implementations. In \cite{Bonzom:2011tf}, these equations have been interpreted as coming from the quantization of a f\/latness constraint and shown to be a reformulation of the f\/lat model for topological BF theory from the Hamiltonian perspective. Projecting the f\/latness constraint on the f\/lux variables, one obtains a Hamiltonian on twisted geometries which is simply the standard relation between the 3d and 4d dihedral angles within a f\/lat
4-simplex. By restriction to the Regge-geometric sector, the Hamiltonian reduces to a constraint introduced in \cite{DittrichRyan}.

The same classical model can be described using the spinor networks introduced before, adding to \eqref{spinornet} a suitable version of the Hamiltonian constraint on a f\/ixed graph, leading to a~spinorial description of~$H_{v,p}$ \cite{Bonzom:2011nv}. The quantum version is then built out of bosonic operators acting on nodes and creating, destroying or exchanging spins~1/2 between two (half-)lines  meeting on a node.

A non-f\/lat case where some contact holds as well concerns symmetry-reduced models.
In the case of the simplest class of non-trivial graphs for spinor networks, i.e. graphs with two vertices~$s$ and~$t$ joined by an arbitrary number $N$ of links, the authors of \cite{Borja:2012xt, Borja:2010gn} def\/ined the dynamics of the 'homogeneous cosmological' sector.  This sector corresponds to the $\U(N)$-invariant sector where there exists a global phase $\phi$ such that $\forall \, l$  $z_l^{s}=e^{i \phi}z_l^{t}$. Its corresponding phase space is a~reduced phase space with two parameters: $\lambda$ represents the total boundary area of the surface separating the two vertices and $\phi$ is its dynamical conjugate variable. $e^{i \phi}$ def\/ines the $\SU(2)$ holonomy living on the edges between the two vertices and $\phi$ thus encodes the curvature. The action def\/ining the dynamics is given by
 \[
S_{\textrm{inv}} [\lambda, \phi] := -2 \int dt \left[ \lambda \partial_t \phi - \lambda^2 ( \gamma^0-\gamma^+ e^{2i \phi} - \gamma^-e^{-2i \phi}) \right],
 \]
 where $\gamma^0$, $\gamma^{\pm}$ are coupling constants. The equations of motion can be solved exactly~\cite{EteraSpinor} and they show that the dynamics can be interpreted as describing homogenous and isotropic cosmology. Moreover, this classical setting is easily quantized and the corresponding Hamiltonian~\cite{Borja:2010gn} shows certain analogies with (the ef\/fective dynamics of) loop quantum cosmology.

\subsection{Covariant theory}

A complete match between discrete Hamiltonians and LQG is obstructed by the fact that solutions to the Hamiltonian constraint are likely to require a superposition of graphs, or an arbitrary f\/ine one. In this optic, we can consider the dynamics on a f\/ixed graph as an approximation, and again postpone the test of dif\/feomorphism symmetry to a later stage, when the way the dynamics changes with the graph comes under control.
Accordingly, we now review the covariant studies of the dynamics. The spacetime manifold is discretized via a simplicial decomposition, or more in general via a cellular decomposition, and approximate the path integral on it. Because we want to use connection variables, the procedure needs an appropriate action.
The more standard procedure is to start with the path integral of general relativity  reformulated as a topological BF gauge theory for the Lorentz group $\SO(3,1)$ (or $\SO(4)$ in the Euclidean case) plus constraints, given by the Plebanski's action\footnote{The Plebanski's action is currently at the heart of the spinfoam models. However, there exists a dif\/ferent approach based on the MacDowell--Mansouri action, which writes general relativity as a BF theory for the gauge group $\SO(4,1)$ (or $\SO(5)$ in the Euclidean case) with a non-trivial potential in the $B$-f\/ield which breaks the symmetry down back to the Lorentz group \cite{FreidelS}. Although this is a very interesting proposition, it has not yet led to a def\/inite proposal for a spinfoam model.}:
\begin{gather} \label{plebanski}
S_{\textrm{Pl}}[B,\omega, \lambda]=\int_\cM  \left( B + \gamma^{-1}\star B\right)^{IJ} \wedge F_{IJ}[\omega]+ \lambda_\alpha \cC_\alpha[B].
\end{gather}
By simplicity's sake, we focus on the Euclidean theory and then $\omega$ is a $\so(4)$-valued 1-form and $F[\omega]$ is its strength tensor, $B$ is a $\so(4)$ valued 2-form and $(\star B)^{IJ}\equiv \f12 \epsilon^{IJ}_{KL}B^{KL}$ is its Hodge dual. The constraints $\cC_\alpha[B]$, enforced by the Lagrange multipliers $\lambda_\alpha$, are  the so-called simplicity constraints. They  constrain the $B$-f\/ield to come from a tetrad f\/ield $e$ in such a way that we recover general relativity in its f\/irst order formalism formulated in term of tetrad and Lorentz connection~\eqref{eq:Holst}. These simplicity constraints turn the non-physical BF theory into 4d gravity. They are second class constraints and modify non-trivially the path integral \cite{AlexandrovBuffenoir, BuffenoirPleb}.

The spinfoam framework is based on a discretized  space-time manifold built from 4-cells glued together. Once again for simplicity's sake and to be in the context of Regge calculus, we consider here only 4d triangulations made of 4-simplices glued together.  We use equivalently  the triangulation $\Delta$ or its dual complex $\Delta^\ast$, the spinfoam 2-complex. Triangles ($\in \Delta$) are dual to faces ($\in \Delta^\ast$) both denoted $f$. Tetrahedra ($\in \Delta$) are dual to links ($\in \Delta^\ast$) both denoted~$t$. 4-simplex are dual to the spinfoam vertices both denoted $v$. %
Both the $B$-f\/ield and the Lorentz connection are discretized. The $B$-f\/ield is a 2-form and is naturally discretized on the triangles $f   \in \Delta$ and the Lorentz connection $\omega$ is discretized as holonomies living on the spinfoam edges~$t$.
Then, the simplicity constraints $\cC_\alpha[B]$ can be discretized: $\cC_\alpha[B] \rightarrow \cC[B_f].$
In all the models constructed, the specif\/ic way the simplicity constraints are discretized and quantized plays a key role.
Two dif\/ferent ways to proceed can be distinguished:
\begin{enumerate}\itemsep=0pt
\item The discretized primary simplicity constraints are turned to quantum operators, acting on the Hilbert space associated with the boundary of each 4-simplex, or 4-cell: $\cC[B_f] \rightarrow \widehat{\cC}[B_f]$. This approach, which can be viewed as a geometrical quantization, is detailed in the next section.
\item The discretized simplicity constraints are included in a discrete constrained BF action. This alternative method is more brief\/ly described in Section~\ref{lagrangian}.
\end{enumerate}

 The (primary) simplicity constraints, $\cC[B_f]$ can be imposed on a given tetrahedron $t$. Then each triangle $f$ of $t$ is characterized by its associated $B_f$-variable. The bivectors satisfy  a~closure constraint $\sum\limits_{f \in t} B^{IJ}_f=0$ as well as the discretized simplicity constraints. As already mentioned in Section~\ref{secCovariantSimplicity}, there exists three dif\/ferent realizations of the simplicity constraints. Their geometrical interpretation is more transparent in the linear formulation: the simplicity constraints come from the fact that all the faces of a given tetrahedron $t$ lay in the same hypersurface. More precisely, in the case with an Immirzi parameter, it is the special combination $(B_f - \gamma \star B_f)^{IJ}$ which has a unique normal for all $f\in t$ and this constraint becomes~\eqref{simplLin} at the quantum level (in fact there will be a sign dif\/ference due to the positive signature of $\Spin(4)$ gauge group).

\subsubsection{Geometrical quantization} \label{secGeoQuant}

Let us now tackle the quantization step and review the dif\/ferent proposals to implement the simplicity constraints at the quantum level.
In this approach, the construction of the so-called spinfoam amplitude is based on the local spinfoam ansatz.
That is the spinfoam amplitude is built from the product of local amplitudes associated to the vertices, edges and faces of the given 2-complex and only depending on the local representations and intertwiners living on those cells. For a given 2-complex $\sigma$ with boundary $\partial \sigma$ and a boundary spin network state $\psi_{\partial \sigma}$, the spinfoam amplitude associated to $\sigma$, $\cA[\sigma,   \psi_{\partial \sigma}]$, consists in a sum over all possible representations and intertwiners living in the bulk and consistent with the boundary spin network:	
 \[
\cA[\sigma,   \psi_{\partial \sigma}]= \sum_{j_f, \, i_{t}} \prod_f \cA_f[j_f] \prod_t \cA_t[i_{t}, j_{f\ni t}] \prod_v \cA_v[j_{f \ni v},i_{t\ni v}],
 \]
where the representations and intertwiners $j_f$ , $i_{t}$ for faces and edges on the boundary $f$ and $t$ $\in \partial \sigma$ are f\/ixed and given by our choice of boundary state $\psi_{\partial \sigma}$.

The key ingredient is the vertex amplitude $\cA_v[j_f, i_t]$ which contains all the dynamical information of the spinfoam model. The proposals reviewed in this section focus solely on the def\/inition of this amplitude.
 It is  computed by the evaluation at the identity of the boundary spinnetwork of the 4-simplex dual to the given vertex $v$. The ef\/fort is to determine the authorized representations and intertwiners to respectively label the edges and nodes of the boundary graph. It is at this stage that the simplicity constraints come into the game directly at the quantum level: they restrict the allowed representations and intertwiners. It is worth  to precise that at this time there is no def\/inite answer on how to implement simplicity constraints which should reintroduce the local degrees of freedom at the quantum level.

 Then usually, the face weight $\cA_f[j_f, i_t]$ is given by the dimension of the representation associated to the given face $f$ whereas there is an ambiguity on the def\/inition of the edge amplitude $\cA_t[j_f, i_t]$.

The quantization procedure for a given tetrahedron $t$, is very simple: an irreducible representation $(j_f^L, j_f^R) \in \N/2\times \N/2$ of the gauge group $\Spin(4)$ is associated to each triangle $t$ and the bivectors $B_f^{IJ}$ are quantized as the $\so(4)$ Lie algebra generators $J^{IJ}_f$ acting in that representation. More precisely in presence of a non-null Immirzi parameter, the symplectic structure on $T^\ast \Spin(4)$ inherited from the Holst--Palatini action is such that the canonical momenta $\Sigma_f$ of the holonomies\footnote{Indeed, to determine the symplectic structure we have to remember that the conf\/iguration variables are the holonomies variables $G_{tt^\prime}$ carried by each link. The link carrying $G_{tt^\prime}$ is the link between the two dual vertices of the tetrahedra $t$ and $t^\prime$. $G_{tt^\prime}$ parallel transports the tetrahedron $t$ to the tetrahedron $t^\prime$.} are deformed by the Immirzi parameter: $\Sigma_f:=B_f+\f1\gamma \star B_f   \leftrightarrow  B_f=\f{\gamma^2}{\gamma^2-1} ( \Sigma -\f1\gamma \star \Sigma_f)$. The quantization step is simply to replace $\Sigma_f^{IJ}$ with the canonical generators of $\spin(4)$, $J_f^{IJ}$.
The closure constraint becomes $\sum\limits_{f \in t} B^{IJ}_f=0  \rightarrow  \sum\limits_{f \in t} J_f^{IJ}=0$ and requires that the quantum states of the tetrahedron are $\Spin(4)$-intertwiners between the representations attached to the tetrahedron triangles.  The primary simplicity constraints become  operators which only involve algebra elements and can be realized in three dif\/ferent ways. These procedure were anticipated in Section~\ref{secCovariantSimplicity}, and we now discuss their implementation at the quantum level. There is also the ``volume part'' of the continuum Lagrangian constraints, which corresponds to secondary constraints in the Hamiltonian analysis, and involve the connection. The strategy to deal with them is to impose a condition of f\/latness of the 4-simplices, which makes them automatically satisf\/ied at the classical level~\cite{EPRlong}.

{\bf The quadratic formulation},
$
\forall\, f, f^\prime \in t$, $\epsilon_{IJKL} J^{IJ}_f J^{KL}_{f^\prime}=0,
$
is the original formulation (without Immirzi parameter). They are the discretized and quantized version of the classical simplicity constraints of the continuum Plebanski's action \eqref{plebanski}.  Working in this section with the Euclidean gauge group $\Spin(4)$, let us recall that they translate into conditions on the Casimir operators on the intertwiners states (see \eqref{simplquadra}). Naturally the f\/irst attempt was to impose strongly all the quantum simplicity constraints. That is to look for intertwiner states $|\psi\ra$ such that:
\begin{gather} \label{quadra}
\epsilon_{IJKL} J^{IJ}_f J^{KL}_{f^\prime}|\psi\ra=0 \qquad \forall\, f,\, f^\prime.
\end{gather}
The spin labels are then constrained to be simple, i.e.\ $j_f^L=j_f^R$ $\forall\, f$. Once the spins $(j_f^L, j_f^R)$ are specif\/ied there exists a unique intertwiner satisfying all constraints \eqref{quadra}, the Barrett--Crane intertwiner $|\psi\ra=i_\textrm{BC}$ \cite{unique}. These restrictions can be implemented  at the level of the partition function\footnote{In particular, considering a single 4-simplex $v$ or its dual boundary graph, Barrett--Crane intertwiners $i_\textrm{BC}$ are attached to each dual vertex (i.e.~tetrahedron) and simple representations to each dual link (i.e.\ triangle). The vertex amplitude is then def\/ined as the evaluation of this boundary spinnetwork at the identity and the obtained result is the $10j$ symbol.} and one gets the well-known Euclidean Barrett--Crane model \cite{BarrettCrane} (for the Lorentzian equivalent model see \cite{BarrettCraneLor}).

However, the uniqueness of the Barrett--Crane intertwiner, $i_\textrm{BC}$ seems to freeze too many degrees of freedom of the 3d space geometry especially when considered from the point of view of Loop Quantum Gravity or from the spinfoam graviton calculations \cite{Alesci:2007tx, Alesci:2007tg, Bianchi:2006uf}. The algebra of these simplicity constraints does not close as a consequence of the non-commutativity of the f\/luxes, see~\eqref{JJJ}. As a consequence, higher and higher order constraints are generated by computing further commutators. That means that by  imposing strongly the quadratic constraints on the intertwiner state $|\psi \ra$, we are actually also imposing all these higher order corrections. Since the uniqueness of the Barrett--Crane intertwiner is a consequence of the imposition of the simplicity constraints, this suggests that the way simplicity constraints are imposed should be modif\/ied.

  To remedy the situation, it was proposed to solve the crossed simplicity constraints weakly, either by using some coherent state techniques \cite{FK, LS2,LS} or by using a Gupta--Bleuler-like method \cite{EPRL, EPRlong, EPR}. The {\it weak} sense means in this context that one requires that $\la \psi |\widehat{\cC}Ê| \phi \ra =0$ for any allowed boundary spinnetwork states. These two approaches were shown to lead to the same spinfoam amplitudes \cite{FK, LS2} for $\gamma<1$, see \cite{Engle:2007mu} for details. The EPRL-FK models rely on the linear reformulation of the simplicity constraints.

{\bf The linear formulation} allows to distinguish a geometrical and a ``non-geometrical" sectors or to introduce very easily the Immirzi parameter in the theory.  In presence of a non-null Immirzi parameter, the expression of the linear simplicity constraints are given by $\cN^I(\star J_f-\gamma^{-1}J_f)_{IJ}=0$ which the Euclidean formulation equivalent to \eqref{simplLin} where the $J$'s are now $\Spin(4)$ generators.
The gauge-f\/ixed version of these constraints $\vec{J}_f-\gamma^{-1}\vec{K}_f=0$ involves in the Riemannian case $\vec{J}_f:=\f12(\vJ_f^L+\vJ_f^R)$ and $\vec{K}_f:=\f12(\vJ_f^L-\vJ_f^R)$.
This formulation is at the root of the construction of the EPRL-FK models. The diagonal simplicity constraints however are f\/irst class, and can be imposed strongly.
In presence of a non-zero Immirzi parameter, they restrict the allowed $\Spin(4)$ representations of each $f$ to $\gamma$-simple representations (up to an ordering ambiguity):
$j_f^L=\rho^2\, j_f^R$, $\forall \, f.
$
In the EPR(L) approach, the linear simplicity constraint~\eqref{defC} is employed to implement weakly the cross-simplicity constraints using a Gupta--Bleuler-like method \cite{EPRL, EPR}. That is, one looks for an Hilbert space~$\cH_s$, subspace of the Hilbert space associated to the tetrahedron $t$, such that the matrix elements of the cross simplicity constraints all vanish.
The strategy is to use the set of linear simplicity constraints \eqref{defC} to form a ``master'' constraint, which selects a subspace of the Hilbert space associated to the tetrahedron, the ``extremum'' subspace: $\cH_s= \bigotimes_{f=1}^4 \cH_{k_f^L + \epsilon k_f^R}$ with $k$ the quantum number associated to the $\SU(2)$ Casimir~$J^2$ and where $\epsilon=+1$ when $\gamma<1$ and $\epsilon=-1$ when $\gamma>1$. Then, the weak imposition of the closure constraint promotes $\cH_s$ to the intertwiner space $\cK_s\equiv \textrm{Inv}_{\SU(2)}[\cH_s]$. Finally, to get a~$\Spin(4)$ spin network, a group averaging on $\Spin(4)$ is performed. The vertex amplitude is obtained as usual by evaluating the boundary spin network of a given 4-simplex labelled with $\gamma$-simple representations and intertwiners taking in $\cK_s$. We get the EPRL vertex amplitude:
\[
\cA_v[j_f, i_t]:=\sum_{\{i_t^L, i_t^R\}} 15j\left(\f{1+\gamma}{2} j_f, i_t^L\right)15j
\left(\f{|1-\gamma|}{2}j_f, i_t^R\right) \bigotimes_{t \subset v} f^{i_t}_{i_t^L, i_t^R},
\]
where the $15j$ are the standard $\SU(2)$ Wigner symbols and $f^{i_t}_{i_t^L, i_t^R}$ are the fusion coef\/f\/icients obtained by contracting $\SU(2)$ intertwiners $i_t$ and $\Spin(4)$ intertwiners $(i_t^L, i_t^R)$ (for further details see~\cite{EPRlong}).

It was also proposed to solve the cross-simplicity constraints weakly by some coherent state method. The aim is the same as the one of the method detailed above. That is the weak imposition of the cross simplicity constraints. The idea proposed in \cite{LS} and developed in \cite{FK, LS2} is to look for semi-classical states such that the simplicity constraints are solved in average, minimizing the uncertainty of these operators. The result obtained is the same as the EPRL one for $\gamma <1$ but it is dif\/ferent for $\gamma>1$. The latter case gives the so-called FK model.

In order to have a geometrical control on the bivectors at the quantum level, the authors of \cite{LS} proposed to work with the following $\SU(2)$ coherent states labeled with an~$\SU(2)$ representation~$j$ and a~unit vector $\hat{n}$: $|j,  \hat{n}\ra\equiv g   |j, j\ra$ with $g\in \SU(2)$, $|j, j\ra$ being the highest weight vector of the standard $\su(2)$ basis. $|j, \hat{n}\ra$ describes in average a 3-vector with norm~$j$ and direction~$\hat{n}$. Its coherence property comes from the fact that the uncertainty is minimal \cite{LS}.

Then, 4-valent\footnote{Since, we are working with a 4d triangulation, we are only interested in 4-valent intertwiner to build boundary spinnetworks. Indeed, the dual of a tetrahedron of $\Delta$ is a 4-valent vertex.} coherent intertwiners are def\/ined by tensoring four such $\SU(2)$ coherent states and group averaging this tensor product in order to get an intertwiner:
\[ 
|| j_f, \hat{n}_f\ra := \int_{\SU(2)} dg \, g \triangleright  \bigotimes_{f=1}^4  |j_f, \hat{n}_f \ra = \int_{\SU(2)} dg\, \bigotimes_{f=1}^4  g  g(\hat{n}_f) |j_f,  j_f\ra,
\]
where the labels are four spins $j_1, \dots, j_4$ and four unit 3-vectors $\hat{n}_1, \dots, \hat{n}_4$. The norm of these intertwiners is peaked on conf\/igurations satisfying the closure constraint $\sum\limits_{f=1}^4j_f \hat{n}_f=0$. Moreover, they form a overcomplete basis of the 4-valent intertwiner space. Then a bivector can be described by the tensor product of $\SU(2)$ coherent states $|j_f^L, \hat{n}_f^L\ra\otimes|j_f^R, \hat{n}^R_f\ra$ where the expectation values of the $\spin(4)$ generators are the two 3-vectors $j^{L,R}\hat{n}^{L,R}$.
The simplicity constraints imply that
\begin{gather}
\f{j^L_f}{j^R_f}=\f{\gamma+1}{|1-\gamma|}.
\label{simplsol}
\end{gather}
Two cases can be distinguished: either $\gamma<1$ and then $\hat{n}^L_f=\hat{n}_f^R$ and we recover the EPRL model, either $\gamma>1$ and then $\hat{n}^L_f=-\hat{n}_f^R$ and we get the FK model. The expression of the vertex amplitude can then be written in terms of the coherent states. This expression in terms of the coherent states allows us to perform its semi-classical analysis (see Section~\ref{semiClassical}).

One drawback with the previous constructions is that the states are not properly def\/ined as actual (strong) solutions of a set of constraints. In particular, they do not come from an actual Gupta--Bleuer procedure with an holomorphic/antiholomorphic factorization of the constraints in term of creation and annihilation operators. This means that we cannot def\/ine the EPRL-FK states through a simple algebraic equation.

 In the contrary, the holomorphic simplicity constraints \eqref{simple_def} do come from such a factorization of the quadratic simplicity constraints \eqref{simplquadra} and allows us to take into account the simplicity constraints by performing a true Gupta--Bleuer procedure.

{\bf The holomorphic formulation} relies on the spinorial framework developed in the context of LQG (see Section~\ref{secCovariantSimplicity}) at the classical level. Working here with the gauge group $\Spin(4)$, a~vertex (before implementation of the simplicity constraints) is characterized by two sets of spinors $\{z_f^{v,L}\}$ and $\{z_f^{v,R}\}$
 satisfying independently the closure constraint. The quantization is straightforward and the spinor components are promoted to annihilation and creation operators
\begin{gather*}
|z_f^{L/R}\ra=\begin{pmatrix}\big(z_f^{L/R}\big)^0\vspace{1mm}\\ \big(z_f^{L/R}\big)^1 \end{pmatrix}   \rightarrow
\begin{pmatrix} a_f^{L/R}\vspace{1mm}\\ b_f^{L/R}\end{pmatrix}, \qquad \la z_f^{L/R} |   \rightarrow   \left(\big(a^{L/R}_f\big)^\dagger,   \big(b^{L/R}_f\big)^\dagger \right),
\end{gather*}
with $[a_f,   a_f^\dagger]=[b_f,  b_f^\dagger]=1$ and $[a_f,  b_f]=0$ for all~$f$.
 The holomorphic simplicity constraint operators for the gauge group $\Spin(4)$ are then
\[
\forall \, e,   f, \qquad a_e^Lb_f^L-a_f^Lb_e^L=\rho^2\big(a_e^Rb_f^R-a_f^Rb_e^R\big),
\]
where we have dropped the index $v$ and $\rho^2:=\f{\gamma+1}{|\gamma -1|}$. These constraints all commute with each other, and can therefore be diagonalized simultaneously. This can be done by means of a~new class of coherent intertwiners, $|\{z_f\}\ra_\rho$, labeled only by the set of spinors $\{ z_f\}$, which have been def\/ined as solution of all the holomorphic simplicity constraints in~\cite{Dupuis:2010iq}. Their relation to the Livine--Speziale coherent intertwiners used in the def\/inition of the EPRL-FK models is explicitly known.

A spinfoam model solving exactly the holomorphic simplicity constraints can be def\/ined \cite{Dupuis:2011dh, EteraHoloEucl}. Its vertex amplitude given by the evaluation of the coherent spin network on the boundary 4-simplex graph obtained by gluing these coherent simple intertwiners $|\{z_f\}\ra_\rho$:
\[
{}_\rho\cA_v(z_f^t)=
\int [dh_t]^5\,e^{\sum\limits_{f\in v}
\rho^2[z_f^{s(f)}|h^L_{s(f)}{}^{-1}h^L_{t(f)}|z_f^{t(f)}\ra
[z_f^{s(f)}|h^R_{s(f)}{}^{-1}h^R_{t(f)}|z_f^{t(f)}\ra}
.
\]
The full spinfoam amplitude is obtained by gluing these vertex amplitudes and integrating over the spinors with a Gaussian measure. Unlike in the EPRL-FK spinfoam model, we do not have simple representations satisfying~\eqref{simplsol}, but rather Gaussian wave-packets peaked on this relation, and all the constraints are treated on the same footing.

\subsubsection{Semi-classical regime} \label{semiClassical}

The contact between these spin foam models and discrete gravity is clear: on each triangulation, or more in general cellular decomposition, the path integral is realized as a sum over histories of discrete 4-geometries associated with the triangulation. Only, these are twisted geometries, represented by areas and angles, and they lack a priori the shape-matching conditions discussed earlier.
It is then remarkable that precisely these conditions are imposed as saddle point equations from the integrals over the group variables. That is, the dominant conf\/igurations in the large spin limit correspond to Regge geometries, correctly weighted by exponentials of the Regge action\footnote{More precisely, by cosines of the Regge actions. Various ways to deal with the presence of both terms have appeared in the literature~\cite{Engle:2011un,Rovelli:2005yj, CarloEdNew}.}.
Therefore we see that, although a precise matching with Regge geometries is lost at the kinematical level, it re-emerges dynamically on each f\/ixed triangulation, providing evidence of the correct semiclassical behaviour of the theory.
The models can be also generalized to a~2-complex of arbitrary valence~\cite{Ding:2010fw, Kaminski:2009fm},
thus providing transition amplitudes for any abstract spin network\footnote{The asymptotic formula, and its relation to Regge calculus,
have still not been studied in this case.}.

Although the relation to Regge calculus on a f\/ixed triangulation is a palatable feature, it is not enough to guarantee the existence of the continuum limit. Much more work is needed to test the formalism.
On the one hand, there are technical details on the def\/inition of the models that still require some thought.
Among these, one that has been often raised in the literature concerns the measure in the path integral. This in turn is related to the specif\/ic structure of the continuum simplicity constraints. As reviewed earlier, there are also secondary simplicity constraints, which are of second class. These are not directly implemented in the EPRL-FK models. Rather, the philosophy is that imposing the primary ones at all times might be suf\/f\/icient.
This is supported by the following fact.
At the canonical level, the secondary constraints ensure that the simplicity of the $B$-f\/ield holds under time evolution. But one can show that if the primary simplicity constraints and the closure constrains are satisf\/ied on each tetrahedron in the boundary of a~\emph{flat} 4-simplex, the secondary simplicity constraints are automatically satisf\/ied \cite{EPR, LS2}.
This is what is done in the EPRL model, where the shape-matching conditions arise as part of the saddle point equations. However, the derivation relies crucially on the f\/latness of the 4-simplex, and at the quantum level, the secondary constraints could undergo non-vanishing f\/luctuations, and the above treatment might fail. An alternative is to implement them as restriction on the group element variables, consistently with their second class nature. This point has been argued by Alexandrov, and more recently in \cite{Alexandrov:2012pj,AlexandrovSigma, Geiller:2011aa}.
Nonetheless, the EPRL construction does lead to a~meaningful and non-trivial restriction of the wave functional dependence on the connection~\cite{Rovelli:2010ed}.

Technical details aside, the key point is the behaviour of the quantum corrections and of possible divergences.
Only a systematic study of higher orders and graph-changing corrections can really test the formalism.

\subsection{Lagrangian methods} \label{lagrangian}

The approach in which one f\/irst quantizes and then imposes the simplicity constraints, has the advantage of leading to a tractable expression, which is manifestly a state sum and a transition amplitude for spin networks, thus realizing the original rationale to introduce the models~\cite{CarloReisenbergerSum}.
On the other hand, away from the large spin semiclassical limit, the action that appears in the path integral does not have an obvious interpretation as a discrete gravity action. For instance, the shape-matching conditions that allow one to recover the Regge action, only appear in the saddle point approximation. An alternative procedure is to insist on a spin foam in which the action is ab initio a discretization of the gravitational action. This approach has been studied in~\cite{Bonzom:2009hw, Bonzom:2008ru, MikeLR}, and more recently in~\cite{Baratin:2011hp}. We refer the interested reader to the review appearing in this same volume, and restrict ourselves to a brief digest of pertinent facts here.

One of the advantages of the spin foam framework is its facility to incorporate multiple varied representations, each of which highlights dif\/ferent facets of the theory as a whole~-- remember that the coherent state representation~\cite{LS2, LS} paved the way to a more transparent geometrical interpretation of the simplicity constraints and catalyzed the development of twisted geometries and the EPRL model.

In an analogous fashion, the non-commutative bi-vector representation~\cite{Baratin:2010nn} allows one to maintain in the discrete theory a remarkable functional similarity to the continuum. It is based on the (non-commutative) Fourier transform existing between a Lie group and its algebra~\cite{Baratin:2010wi},  For a theory in the mould of BF theory plus constraints (Plebanski), this entails a transformation from holonomies ($G_f$) to bi-vectors~($B_f$).  With this in hand, one can impose the closure and Lagrangian simplicity constraints directly upon the bi-vectors, which is how they are implemented in the continuum. As we have stressed already, in passing from the continuum to a discrete setting, one has a choice in how one discretizes the constraints.  Interestingly, the interplay of closure and simplicity suggests a form of the simplicity constraints that leads to a class of spin foam models similar but dif\/ferent from those considered so far.  The claim is that in imposing the Lagrangian simplicity constraints in this fashion (and as usual, within every simplex of the discrete manifold), one is implicitly, yet more completely, capturing the secondary simplicity constraints (of the discrete canonical theory).  That is to say, this model puts these constraints into ef\/fect in the quantum theory rather than as saddle point equations in the semi-classical regime.  These proposals are under active investigation.

 \section{Conclusions} \label{secConcl}

The last few years have seen a number of interesting developments in Loop Quantum Gravity, based on taking seriously an interpretation in terms of discrete geometries of the truncation of the theory to a f\/ixed graph.
This interpretation becomes particularly useful in the study of certain spin foam models, notably the EPRL model, where the large spin limit is dominated precisely by exponentials of the Regge action.
The interpretation has helped sheding light on the use of coherent states, on the def\/inition and implementation of the simplicity constraints, and brought to surface a number of intriguing new ideas, such as spinor and twistor tools \cite{IoZako}, and ${\rm U}(N)$ symmetries \cite{Borja:2011pd}. The hope is that some of these ideas and tools can also help show the way to understanding the complete dynamics of the theory, beyond the single graph truncation.

\pdfbookmark[1]{References}{ref}
\LastPageEnding

\end{document}